\documentclass[journal]{IEEEtran}
\IEEEoverridecommandlockouts

\usepackage{stmaryrd}
\usepackage{amsmath}
\usepackage{amsthm}
\usepackage{xfrac}
\usepackage{amssymb}
\usepackage{mathabx}
\usepackage{psfrag}
\usepackage{graphicx}
\usepackage{trimclip}
\usepackage{epstopdf}
\usepackage{todonotes}
\usepackage{cite}
\usepackage{algorithmic}
\usepackage{algorithm}
\usepackage{svg}
\usepackage{mathtools}
\usepackage{hyphenat}
\usepackage{booktabs}
\usepackage{hyperref}
\usepackage{multirow}
\usepackage{multicol}
\usepackage[caption=false]{subfig}

\usepackage{amsmath,amsfonts,amssymb}
\usepackage{algorithmic}
\usepackage{array}

\usepackage{cite}
\usepackage{amsmath,amssymb,amsfonts}
\usepackage{algorithmic}
\usepackage{graphicx}
\usepackage{textcomp}
\usepackage{xcolor}

\usepackage{accents}
\usepackage{minibox}
\usepackage{fancyhdr}
\usepackage{kantlipsum}
\fancypagestyle{fancy}

\fancypagestyle{plain}{
  \fancyhf{}
  \fancyhead[C]{Conference on \LaTeX}     %% C or L or R.
  \fancyfoot[L]{This is a notice}%                        %% C or L or R.

}
\usepackage{eso-pic}

\def\BibTeX{{\rm B\kern-.05em{\sc i\kern-.025em b}\kern-.08em
    T\kern-.1667em\lower.7ex\hbox{E}\kern-.125emX}}

\usepackage{algorithm}
\usepackage{algorithmic}
\usepackage{amsthm}

\DeclareMathOperator*{\argmin}{arg\,min}
\DeclareMathOperator{\diag}{diag}
\DeclareMathOperator{\blkdiag}{blkdiag}
\DeclareMathOperator{\myvec}{vec}
\DeclareMathOperator{\myspan}{span}
\DeclareMathOperator{\rank}{rank}
\DeclareMathOperator{\myproj}{proj}

\DeclareMathOperator*{\Tr}{Tr}

\newtheorem{thm}{Theorem}
\newtheorem{lemma}{Lemma}

\newcommand{\mtrian}{\mathrel{\raisebox{-0.1ex}{%
\scalebox{0.8}[0.6]{$\vartriangle$}}}}

\begin{document}

% Add IEEE pre-print notice
\AddToShipoutPictureBG*{%
  \AtPageUpperLeft{%
    \hspace*{\dimexpr0.175\paperwidth\relax}%%  change \dimexpr0.5\paperwidth\relax appropriately
    % \chead{\large This work has been submitted to the IEEE for possible publication. Copyright may be transferred without notice, after which this version may no longer be accessible.}
    % \makebox(0,-0.75)[c]{\large This work has been submitted to the IEEE for possible publication. Copyright may be transferred without notice, after which this version may no longer be accessible.}%
    \minibox[c]{\\ \\ \\ \\ \emph{This work has been submitted to the IEEE for possible publication.} \\ \emph{Copyright may be transferred without notice, after which this version may no longer be accessible.}}
}}
\AddToShipoutPictureBG*{%
  \AtPageLowerLeft{%
    \setlength\unitlength{1in}%
    \hspace*{\dimexpr0.5\paperwidth\relax}%%  change \dimexpr0.5\paperwidth\relax appropriately
    \makebox(0,0.75)[c]%
}}

\title{{Towards xAI: Configuring RNN Weights using Domain Knowledge for MIMO Receive Processing}}
\author{Shashank Jere, Lizhong Zheng, Karim Said and Lingjia Liu
\thanks{Part of this work was presented in~\cite{JereMILCOM2023} at the \emph{2023 Military Communications Conference} (MILCOM'23) in Boston, MA, USA. S. Jere, K. Said and L. Liu are with \emph{Wireless@VT}, Bradley Department of Electrical and Computer Engineering at Virginia Tech. L. Zheng is with the EECS Department at Massachusetts Institute of Technology. The corresponding author is L. Liu (ljliu@ieee.org).}}

\maketitle

\begin{abstract}

Deep learning is making a profound impact in the physical layer of wireless communications. Despite exhibiting outstanding empirical performance in tasks such as MIMO receive processing, the reasons behind the demonstrated superior performance improvement remain largely unclear. 
In this work, we advance the field of Explainable AI (xAI) in the physical layer of wireless communications utilizing signal processing principles.
Specifically, we focus on the task of MIMO-OFDM receive processing (e.g., symbol detection) using reservoir computing (RC), a framework within recurrent neural networks (RNNs), which outperforms both conventional and other learning-based MIMO detectors.
Our analysis provides a signal processing-based, first-principles understanding of the corresponding operation of the RC.
Building on this fundamental understanding, we are able to systematically incorporate the domain knowledge of wireless systems (e.g., channel statistics) into the design of the underlying RNN by directly configuring the untrained RNN weights for MIMO-OFDM symbol detection.
The introduced RNN weight configuration has been validated through extensive simulations demonstrating significant performance improvements.
This establishes a foundation for explainable RC-based architectures in MIMO-OFDM receive processing and provides a roadmap for incorporating domain knowledge into the design of neural networks for NextG systems.

\end{abstract}


\section{Introduction}
\label{sec:Introduction}
The recent surge in deep learning has been extraordinary, largely due to its great empirical success across a broad range of applications. 
Wireless communications have also embraced machine learning (ML) and neural network (NN) techniques at a rapid pace.
Artificial intelligence (AI)-aided networks have been foreseen to play a crucial role in addressing the stringent requirements and challenges posed by next-generation  wireless networks (NextG)~\cite{Shafin2020, Xu2024}.
Specifically, NextG brings forth the challenges~\cite{Shafin2020} of \emph{network complexity}, \emph{model deficit} and \emph{algorithm deficit}, thereby limiting the feasibility of traditional model-based approaches for physical layer (PHY) processing, including MIMO receive processing.  
Meanwhile, AI-enabled methods for PHY processing can offer an attractive solution to overcome these challenges~\cite{Xu2024}.  
Most prevailing deep learning approaches entail training large NN models ``offline'' with large datasets before deploying them for inference, which may not be practical for many wireless operations, particularly at the PHY where over-the-air (OTA) training data is scarce. 
Another formidable challenge in the application of learning-based methods at the PHY is the \emph{uncertainty in generalization}~\cite{Shafin2020}, caused primarily due to mismatch between offline training and online deployment environments in terms of system configurations, channel environments, and operational adaptations~\cite{Xu2024}.
This makes the direct application of purely offline learning methods infeasible for training data-starved low-latency PHY applications.
Alternatively, online learning and hybrid learning approaches have been viewed as pathways towards addressing the above mentioned challenges. 
Toward this end, the reservoir computing (RC) paradigm~\cite{Lukosevicius2012} enables online real-time learning strategies owing to its low-complexity training methodology, making it ideal for complexity-constrained and latency-aware PHY operations such as MIMO-OFDM receive processing~\cite{zhou2019,zhou2020rcnet,RCstruct} and dynamic spectrum access~\cite{LiuAI2,Chang2020}, demonstrating superior performance compared to conventional model-based methods and other offline learning approaches~\cite{Khani2020,DSAComparison}.

Deep neural network (DNN)-based MIMO symbol detection methods have gained large traction recently, leading to significant progress. 
Multi-layer perceptron (MLP)-based detection techniques have been introduced in works such as \emph{DetNet}~\cite{samuel2019learning}, \emph{MMNet}~\cite{Khani2020}, \emph{OAMPNet}~\cite{OAMPNet2018}, and \emph{HyperMIMO}~\cite{Goutay2020}, where each approach integrates trainable parameters from traditional iterative algorithms. 
While these methods show promising performance, they often require large amounts of training data, making them difficult to implement in modern cellular networks such as 5G NR and 5G-Advanced, where the available training data in the form of OTA reference signals (RS) is extremely limited.
Moreover, they typically rely on perfect channel state information (CSI), which is challenging, if not impossible, to obtain in practice.
Given the lightweight training requirements of reservoir computing (RC)-based techniques, they offer a promising alternative.
Within the RC framework, echo state networks (ESNs) were first applied for symbol detection in MIMO-OFDM systems in~\cite{mosleh2017brain}. Enhancements to the ESN architecture, such as the ability to process `windowed' inputs and additional output skip and delay connections, were introduced in~\cite{zhou2019}.
The new architecture is called \emph{WESN}, demonstrating significant performance gains over vanilla ESN. 
Further advancements include the introduction of the deep RC structure \emph{RCNet}~\cite{zhou2020rcnet}, and \emph{RC-Struct}~\cite{RCstruct}, which leverages the time-frequency structure of OFDM waveforms. 
Both approaches showed substantial improvements over traditional signal processing methods and other learning-based approaches.
A key advantage of such RC-based methods over other established NN-based MIMO detectors is that their training is fully online and can be done in a slot-basis with significantly lower computational complexity. This makes RC-based approaches more resilient to slot-based dynamic transmission mode adaptation in modern cellular networks since the detector is retrained in real-time within each new slot using the limited OTA RSs.
Despite the empirical evidence, a systematic and theoretical analysis of the general effectiveness of RC-based methods and the explainability of their superior performance in PHY receive processing have not been fully developed yet.

In parallel with the rise of deep learning methods, there has been a steady growth in the need for explainability of such methods, leading to the rise of the field of explainable artificial intelligence (xAI) and explainable machine learning (xML)~\cite{Montavon2019}. 
One of the first works that explore the theoretical underpinnings of the success of RC in time-series problems is~\cite{Ozturk2007}, which introduces a functional space approximation framework.
Another noteworthy recent work~\cite{Bollt2021} shows that an ESN with linear activation is equivalent to performing vector autoregression (VAR). 
The ability of RC to predict complex nonlinear dynamical systems, including the Lorenz and Rössler systems, was examined in~\cite{Halus2019}, whereas~\cite{Carroll2022} explored the tuning and optimization of the fading memory length in RC systems.
In our earlier work~\cite{JereTCOM2023}, we established an upper bound on the Empirical Rademacher Complexity for single-reservoir ESNs, demonstrating that ESNs offer tighter generalization compared to traditional RNNs. Additionally, we highlighted the practical value of the derived bound in optimizing an ESN-based symbol detector for MIMO-OFDM systems.
Other works grounded in statistical learning theory, such as~\cite{Gonon2020}, also derive bounds for the generalization error of RC by utilizing modified versions of Rademacher-type complexity measures.
In our prior work~\cite{Jere2023WCL}, we provided a signal processing analysis of the ESN and offered a comprehensive analytical characterization of the optimal untrained recurrent weight for a single-neuron ESN under the task of channel equalization.
Although existing literature offers some valuable insights, a clear signal processing perspective combined with comprehensive analytical characterizations for RC-based techniques has yet to be fully developed.
Our current work aims to bridge this gap in the context of RC-based symbol detection for MIMO-OFDM systems, building on and extending our previous works in this domain.
The insights gained from this work can be potentially utilized to enhance the explainability of general NN-based approaches for various engineering applications.

Our contributions in this work are summarized below:
\begin{itemize}
    \item Building on our previous work~\cite{JereMILCOM2023}, we introduce a new ``time-domain'' approach for configuring the untrained weights of the ESN for strictly minimum-phase channels, extending the frequency-domain approach of untrained weight configuration introduced in~\cite{JereMILCOM2023}.

    \item We provide a theoretical explanation of the linkage between symbol detection over a non-minimum-phase wireless channel and the WESN structure introduced in our earlier work~\cite{zhou2019}. This theoretically justifies the previously introduced WESN architecture and introduces the explainability of the architecture which is at the heart of RC-based detectors adopted in our subsequent works.

    \item We extend the frequency and time-domain approaches for untrained weight configuration of the WESN for both OFDM and MIMO-OFDM symbol detection utilizing the parametric MIMO channel representation.

    \item We validate the OFDM and MIMO-OFDM WESN weight configuration procedures through extensive simulations under 5G/5G-Advanced scenarios.
\end{itemize}

\emph{Notation:} 
Zero-based indexing is used in this paper.
$(*)$ denotes the linear convolution operation.
$\mathbf{a}$ denotes a column vector. $\boldsymbol{A}$ and $\mathbf{A}$ denote matrices.
$[\mathbf{A}]_{:, i:j}$ denotes the submatrix containing columns $i$ through $j$ of $\mathbf{A}$.
$[\cdot, \cdot]$ denotes horizontal concatenation and $[\cdot ; \cdot]$ denotes vertical concatenation respectively of scalars, vectors or matrices.
$(\cdot)^*$ denotes the complex conjugate. 
$\boldsymbol{0}_{M \times N}$ denotes the $M \times N$ all-zeros matrix.
$\mathbf{I}_N$ denotes the $N \times N$ identity matrix.
$(\cdot)^H$ denotes the matrix conjugate transpose.
$(\cdot)^{\dagger}$ denotes the Moore-Penrose matrix pseudoinverse.
$\Tr(\cdot)$ denotes the matrix trace.
$\mathcal{T}(\mathbf{a})$  produces a lower triangular Toeplitz matrix with $\mathbf{a}$ as its first column.
$\diag(\cdot)$ denotes a diagonal matrix formed by scalar arguments; 
$\blkdiag(\cdot)$ denotes a block diagonal matrix formed by matrix/vector arguments.
We use the abbreviations `MP' for `minimum-phase' and `NMP' for `non-minimum-phase'.

The rest of the paper is organized as follows. 
Sec.~\ref{sec:rc_preliminaries_and_system_model} provides preliminaries on RC and introduces the time-domain weight configuration formulation for OFDM symbol detection.
Sec.~\ref{sec:rc_design_objective} provides a systematic way of configuring the RNN weights of RC using the domain knowledge.
Sec.~\ref{sec:nmp_channels_wesn_connection} sets up the theoretical foundation to analyze NMP channels and its connection to the WESN, providing a theoretical justification for its architecture.
Sec.~\ref{sec:mimo_equalization} extends the introduced analysis and approach to MIMO-OFDM symbol detection utilizing the parametric MIMO channel representation. Sec.~\ref{sec:performance_evaluation} validates the introduced theoretical analysis and weight configuration strategies via extensive simulations.  Finally, Sec.~\ref{sec:conclusion} concludes the paper.

\section{RC-based OFDM Symbol Detection}
\label{sec:rc_preliminaries_and_system_model}
\subsection{The Vanilla Echo State Network}
\label{sec:conventional_esn}
\begin{figure}[htbp]
    \centering    \includegraphics[width=0.65\linewidth]{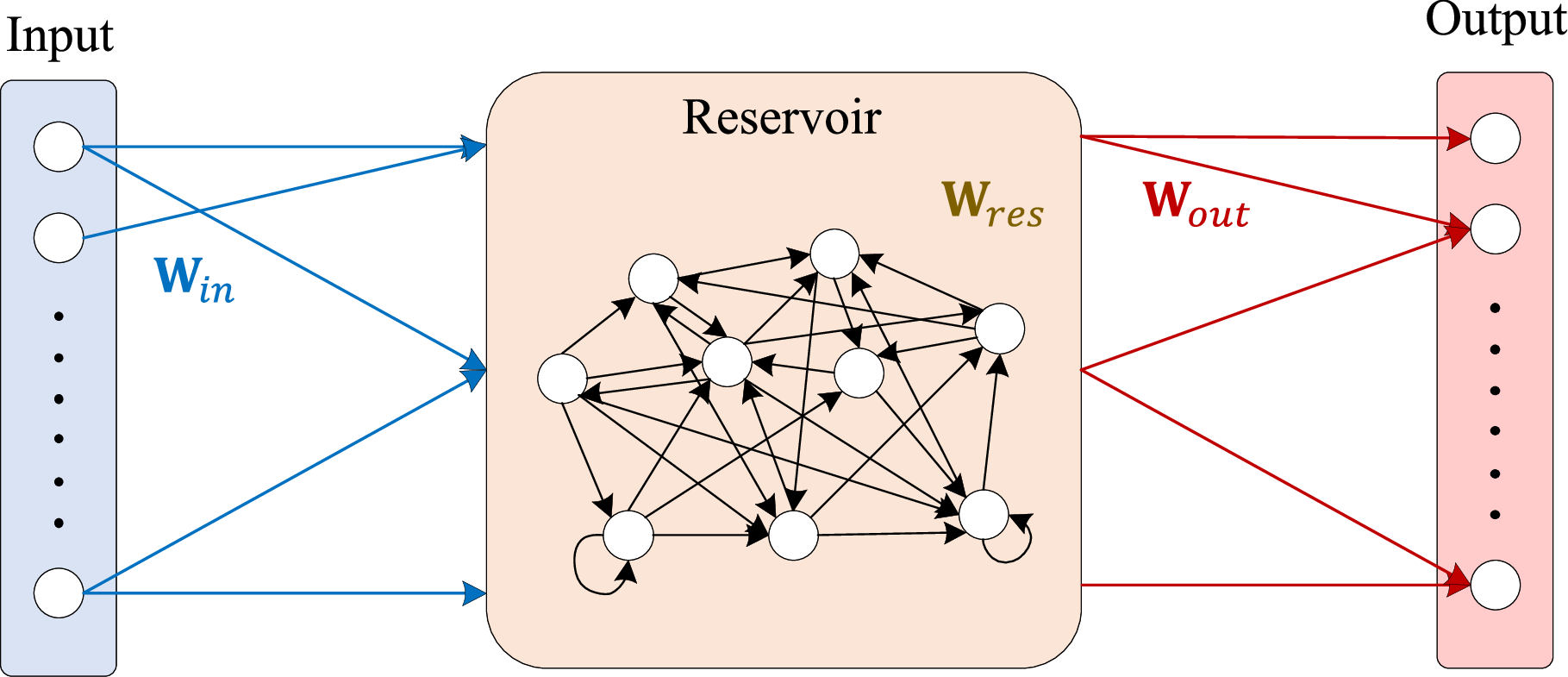}
    \caption{A single reservoir `vanilla' ESN}
    \label{fig:ESN_figure}
\end{figure}

ESN, a popular architecture within the RC paradigm, comprises a ``reservoir''/RNN containing $N_{\mathrm{n}}$ randomly inter-connected neurons, along with an input weights layer and an output (readout) weights layer with the following parameters:
\begin{itemize}
    \item $\mathbf{x}_{\mathrm{in}}[n] \in \mathbb{C}^{d_{\mathrm{in}}}$: Input to ESN at time index $n$.
    \item $\mathbf{x}_{\mathrm{res}}[n] \in \mathbb{C}^{N_{\mathrm{n}}}$: Reservoir state vector at time index $n$. 
    \item $\mathbf{W}_{\mathrm{in}} \in \mathbb{C}^{N_{\mathrm{n}} \times d_{\mathrm{in}}}$: Input weights matrix.
    \item $\mathbf{W}_{\mathrm{res}} \in \mathbb{C}^{N_{\mathrm{n}} \times N_{\mathrm{n}}}$: Reservoir weights matrix.
    \item $\mathbf{W}_{\mathrm{out}} \in \mathbb{C}^{d_{\mathrm{out}} \times N_{\mathrm{n}}}$: Output weights matrix.
    \item $\mathbf{x}_{\mathrm{out}}[n] \in \mathbb{C}^{d_{\mathrm{out}}}$: Output of ESN at time index $n$. 
\end{itemize}
Here, $d_{\mathrm{in}}$/$d_{\mathrm{out}}$ is the input/output dimension of the ESN representing the number of features in the input/output.
With a nonlinear activation function $\sigma(\cdot)$ (e.g., ReLU, tanh, etc.), the state update and the output equations can be expressed as:
\begin{align}
\mathbf{x}_{\mathrm{res}}[n] &= \sigma \big(\mathbf{W}_{\mathrm{res}}\mathbf{x}_{\mathrm{res}}[n-1] + \mathbf{W}_{\mathrm{in}}\mathbf{x}_{\mathrm{in}}[n] \big)
\label{eq:state_transition_eqn}, \\
\mathbf{x}_{\mathrm{out}}[n] &= \mathbf{W}_{\mathrm{out}}\mathbf{x}_{\mathrm{res}}[n].
\label{eq:output_eqn}
\end{align}
A vanilla ESN with a single reservoir is shown in Fig.~\ref{fig:ESN_figure}.
For analytical tractability, we consider the ``\textbf{linear}'' ESN with $\sigma(\cdot)$ being the identity function. 
This allows analyzing the effectiveness of RC architectures in a tractable manner, similar to~\cite{Bollt2021}.
In the RC convention, only $\mathbf{W}_{\mathrm{out}}$ is trained using a least-square approach, whereas $\mathbf{W}_{\mathrm{in}}$ and $\mathbf{W}_{\mathrm{res}}$ are randomly generated from a particular pre-determined distribution (e.g., uniform distribution) and then kept fixed throughout the training and the test (inference) stages.
Specifically for the vanilla ESN, $\mathbf{W}_{\mathrm{out}}$ is found as $\widehat{\mathbf{W}}_{\mathrm{out}} = \argmin_{\mathbf{W}_{\mathrm{out}}} \| \mathbf{W}_{\mathrm{out}} \mathbf{S} - \boldsymbol{O}  \|_F^2 = \boldsymbol{O} \mathbf{S}^{\dagger}$, where $\mathbf{s}[n] \overset{\mtrian}{=} \mathbf{x}_{\mathrm{res}}[n]$, $\mathbf{S} = \left[\mathbf{s}[0], \mathbf{s}[1], \ldots, \mathbf{s}[T-1] \right] \in \mathbb{C}^{N_{\mathrm{n}} \times T}$ is the concatenated reservoir states matrix for input and output sequences of length $T$ and $\boldsymbol{O} \in \mathbb{C}^{d_{\mathrm{out}} \times T}$ is the training label. 
Our hypothesis is that the impact of nonlinear activation in the reservoir and the configuration of $\mathbf{W}_{\mathrm{in}}$ and $\mathbf{W}_{\mathrm{res}}$ are either orthogonal or at least separable, as evidenced empirically in our recent works~\cite{Jere2023WCL, JereMILCOM2023}.

\subsection{Signal Processing Preliminaries for the Vanilla ESN}
\begin{figure}[htbp]
    \centering    \includegraphics[width=0.675\linewidth]{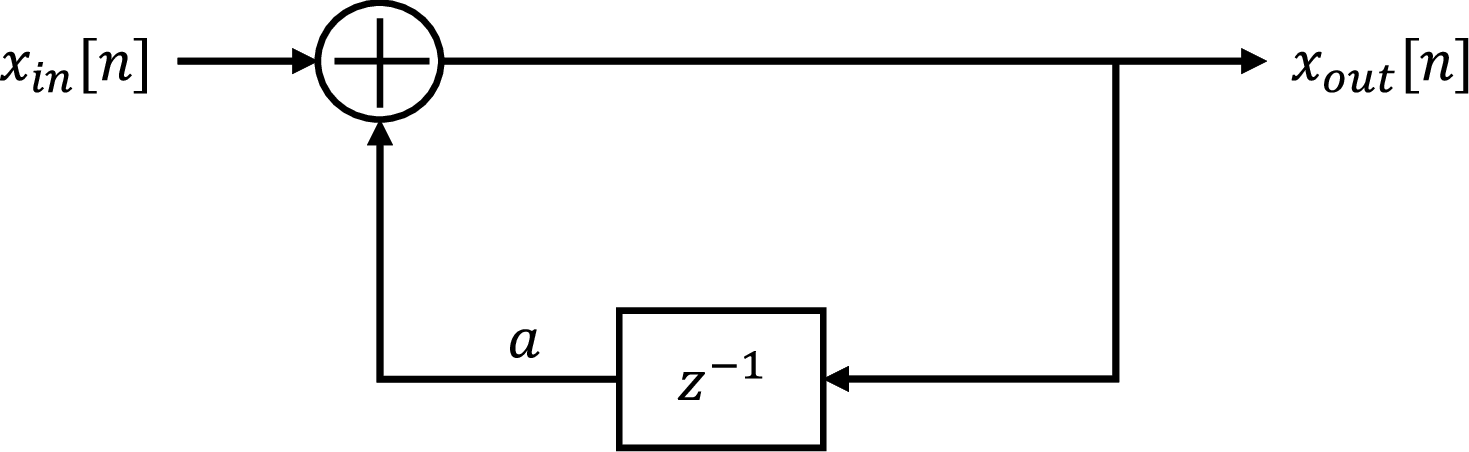}
    \caption{Modeling a neuron in the reservoir as a single-pole IIR filter.}
    \label{fig:single_pole_iir}
\end{figure}

Based on our previous work~\cite{Jere2023WCL}, we know that the vanilla ESN with a single recurrent neuron and linear activation is modeled as a first-order (single-pole) infinite impulse response (IIR) filter.
The block diagram for this structure is shown in Fig.~\ref{fig:single_pole_iir}.
The system response (transfer functions) of this filter is given by $H_{0}(z) = \sfrac{X_{\mathrm{out}}(z)}{X_{\mathrm{in}}(z)} = \frac{1}{1-az^{-1}}$.
In general, the ESN reservoir with $N$ recurrent neurons can be modeled as an IIR filter whose transfer function is written as $H_{\mathrm{res}}(z) = \frac{\sum_{k=0}^M b_k z^{-k}}{1 + \sum_{k=1}^N a_k z^{-k}}$, where $M \leq N$.
For a given realization of a wireless channel with system response $H_{\mathrm{ch}}(z)$, the objective of the ESN-based equalizer is to learn the inverse mapping $\widehat{H}_{\mathrm{inv}}(z)$ from the available training data while minimizing the residual error, so that $\widehat{H}_{\mathrm{inv}}(z) H_{\mathrm{ch}}(z) \approx 1$.
In the OFDM context, the ESN applied as a time-domain equalizer aims to recover the transmitted time-domain OFDM symbol directly using embedded training data (e.g., RS), thus completely bypassing traditional channel estimation.
For a strictly MP channel, the direct inverse given by $\widehat{H}_{\mathrm{inv}}(z) = 1/H_{\mathrm{ch}}(z)$ is stable and the ESN-based equalizer can attempt to learn either the corresponding frequency-domain inverse $1/H_{\mathrm{ch}}(e^{j\omega})$ or the time-domain equalizer impulse response $\mathbf{g}$ such that $\mathbf{g} * \mathbf{h} = \delta[n]$, where $\mathbf{h} \in \mathbb{C}^L$ is the time-domain impulse response of the channel with the system function $H_{\mathrm{ch}}(z)$.  
We extend this analysis to NMP and mixed-phase channels in Sec.~\ref{sec:nmp_channels_wesn_connection}, where the role of feedforward taps of a finite impulse response (FIR) filter is highlighted.

\subsection{Geometric Interpretation of the Vanilla ESN}
\begin{figure}[h]
    \centering    \includegraphics[width=0.675\linewidth]{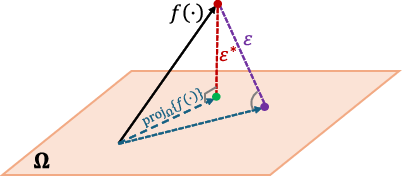}
    \caption{A geometric interpretation of training of output weights in a vanilla ESN as an orthogonal projection.}
    \label{fig:esn_geometric_interpretation}
\end{figure}

Our previous works~\cite{JereMILCOM2023, JereJSTSP2024} have introduced the ``geometric interpretation'' of the vanilla ESN.
In this framework, we consider an ESN with a reservoir consisting of $K$ parallel recurrent neurons that are \emph{not} connected to each other.
Given that the transfer function of the $k^{th}$ neuron is $H_{k}(z)$, the constituent neurons in the reservoir span a subspace $\boldsymbol{\Omega}$ such that $\boldsymbol{\Omega} = \myspan{\{ H_{k}(z)} \}_{k=1}^{K}$.
For a target function $f(\cdot)$, the approximation of $f(\cdot)$ generated by the ESN is a linear combination of the basis functions $\{H_{k}(z) \}$, in general with an approximation error of $\varepsilon$. 
Considering $\ell_{2}$-norm as the loss function used to train the combining weights represented by the output weights matrix $\mathbf{W}_{\mathrm{out}}$, the ESN's approximation to $f(\cdot)$ becomes an orthogonal projection of $f(\cdot)$ onto $\Omega$, which can be denoted as $f_{\perp}(\cdot) = \myproj_{\Omega}{\{ f(\cdot)} \}$.
This orthogonal projection achieves an approximation error of $\varepsilon^*$, the minimum possible $\ell_2$-loss.  
In addition, the choice of the $\ell_{2}$-norm as the loss function transforms the training of $\mathbf{W}_{\mathrm{out}}$ into a simple least-square problem.
Since this training methodology does not require backpropagation-based training, it greatly reduces the training computational complexity of the ESN, thereby admitting real-time PHY operations such as symbol detection.
Here, the target function $f(\cdot)$ can be completely arbitrary and its knowledge is obtained only through the training samples.
\emph{Therefore, the geometric interpretation of the ESN is applicable to a wide range of learning problems in engineering applications.}
In the context of symbol detection, $f(\cdot)$ can be the frequency-domain response of the channel inverse or the impulse response of the time-domain equalizer.
The geometric interpretation of the vanilla ESN described above is visualized in Fig.~\ref{fig:esn_geometric_interpretation}.

\subsection{Frequency-Domain View of RC-based Equalization}
\label{sec}
\begin{figure}[h]
    \centering    \includegraphics[width=\linewidth]{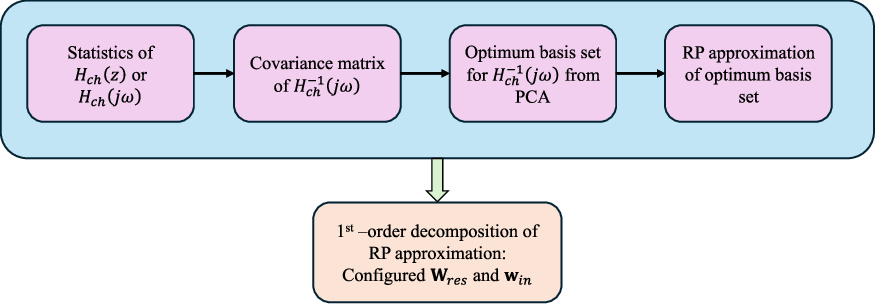}
    \caption{Summary of the frequency-domain reservoir/RNN weight configuration procedure of the vanilla ESN in~\cite{JereMILCOM2023}.}  \label{fig:freq_domain_method_flowchart}
\end{figure}

In our previous work~\cite{JereMILCOM2023}, we considered the case of configuring the untrained weights of the vanilla ESN for the symbol detection task, given the statistics of the underlying wireless channel.
To this end, a `frequency-domain' approach was introduced.
The overall OFDM symbol detection task was divided into ESN-based time-domain equalization of the received signal, followed by classification in the frequency domain.
Furthermore, we had considered strictly MP channels, i.e., if the channel impulse response in the transform domain ($z$-domain) is denoted by $H_{\mathrm{ch}}(z)$, then the roots of $H_{\mathrm{ch}}(z)$ lie strictly inside the unit circle in the complex plane.
To summarize, a systematic procedure was introduced to obtain the configured weights $\mathbf{W}_{\mathrm{in}}$ and $\mathbf{W}_{\mathrm{res}}$ of the vanilla ESN.
Specifically, we described a method that uses the empirically collected channel statistics in the frequency-domain followed by principal component analysis (PCA), and a rational polynomial (RP) approximation of the basis vectors to obtain the configured weights.
A complete description of this procedure together with performance evaluation in strictly MP channels was provided in~\cite{JereMILCOM2023}.
A summary of the frequency-domain configuration procedure is illustrated in Fig.~\ref{fig:freq_domain_method_flowchart}.
In this work, we build on the frequency-domain approach and introduce a time-domain perspective as well as a systematic procedure for configuring the untrained weights of the vanilla ESN which also extends to the WESN introduced in~\cite{zhou2019}.

\subsection{Time-Domain View of RC-based Equalization}
Consider a discrete-time representation of a wireless channel with an impulse response given by $\mathbf{h}=[h_0, h_1, \ldots, h_{L-1}]^T \in \mathbb{C}^{L} $.
The corresponding system response is given by $ H_{\mathrm{ch}}(z) = \sum_{\ell=0}^{L-1} h_{\ell}z^{-\ell}$. 
Consider an input sequence $\mathbf{x} \in \mathbb{C}^N$ transmitted through the channel with the above impulse response.
The received sequence obtained via linear convolution is given by $\mathbf{y} = \mathbf{h} * \mathbf{x} + \mathbf{n}$,
where $\mathbf{n} \sim \mathcal{CN}(\boldsymbol{0}, \sigma^2 \mathbf{I}_N)$ is the additive white Gaussian noise (AWGN) with variance $\sigma^2$.
This operation is equivalently represented as $\mathbf{y} = \widetilde{\mathbf{H}} \mathbf{x} + \mathbf{n}$,
where $\widetilde{\mathbf{H}} \in \mathbb{C}^{N \times N}$ is a lower triangular Toeplitz convolution matrix given by $\widetilde{\mathbf{H}} = \mathcal{T}([\mathbf{h}^T, \boldsymbol{0}_{(N-L) \times 1}^T]^T)$.
The objective of a learning-based equalizer is to find an inverse mapping $\widetilde{\mathbf{G}} \in \mathbb{C}^{N \times N}$ such that $\mathbb{E}_{\mathbf{h}} [\| \widehat{\mathbf{x}} - \mathbf{x} \|_2^2]$ is minimized, where $\widehat{\mathbf{x}} = \widetilde{\mathbf{G}} \mathbf{y}$ and $\widetilde{\mathbf{G}} = \widetilde{\mathbf{H}}^{-1}$ is also a lower triangular Toeplitz matrix and can be written as $\widetilde{\mathbf{G}} = \mathcal{T}(\mathbf{g})$, where $\mathbf{g} \overset{\mtrian}{=} [g_0, \; g_1, \; \ldots, \; g_{N-1}]^T = [\widetilde{\mathbf{G}}]_{:,0} \in \mathbb{C}^{N}$ denotes the impulse response of the ideal zero-forcing (ZF) equalizer for the channel $\mathbf{h}$.
The above formulation serves as the foundation for the time-domain weight configuration procedure described next in Sec.~\ref{sec:optimum_basis_pca} and Sec.~\ref{sec:optimum_basis_to_weights}.

\subsection{Optimum Orthogonal Basis Set in the Time-Domain}
\label{sec:optimum_basis_pca}
Given a time-domain impulse response $\mathbf{h}$ with the corresponding convolution matrix $\widetilde{\mathbf{H}}$ for a particular channel realization, the Toeplitz form of the corresponding ZF equalizer is $\widetilde{\mathbf{G}} = \widetilde{\mathbf{H}}^{-1}$.
Thus, for the $i^{\text{th}}$ strictly MP channel realization denoted as $\mathbf{h}_{(i)}$, the impulse response of its equalizer is $\mathbf{g}_{(i)}$, where $\mathbf{g}_{(i)} = [\widetilde{\mathbf{G}}_{(i)}]_{:, 0}$. 
In order to find the optimum basis set for the subspace spanned by $\{ \mathbf{g}_{(i)} \}$, we employ PCA, similar to the approach adopted in~\cite{JereMILCOM2023}.
For $N_{\mathrm{obs}}$ realizations of such equalizer impulse responses, the empirical covariance matrix can be computed as $\widehat{\mathbf{K}}_{gg} = \frac{1}{N_{obs}} \sum_{i=1}^{N_{obs}} \mathbf{g}_{(i)} \mathbf{g}_{(i)}^H$, whose eigen-decomposition is obtained as $\widehat{\mathbf{K}}_{gg} = \mathbf{V} \mathbf{\Lambda} \mathbf{V}^H$, where the columns of $\mathbf{V} \in \mathbb{C}^{N \times N}$ contain the  eigenvectors and $\mathbf{\Lambda} = \diag(\lambda_0, \ldots, \lambda_{N-1})$ contains the corresponding eigenvalues.
Finally, the set of $M$ optimum basis vectors for the subspace spanned by $\{ \mathbf{g}_{(i)} \}$ can be collected in the columns of $\mathbf{F} \in \mathbb{C}^{N \times M}$ as simply the set of the first $M$ columns of $\mathbf{V}$, i.e., $\mathbf{F} \overset{\mtrian}{=} [\mathbf{V}]_{:, 0:M-1}$.
It can be seen that except the nature of the inverse function (frequency response versus impulse response), the procedure to find the optimum basis matrix $\mathbf{F}$ is identical in the time-domain method compared to the frequency-domain method introduced in~\cite{JereMILCOM2023}. 

\textbf{Minimum-Phase Basis:}
Despite the similarity between the frequency-domain and the time-domain procedures of finding the optimum basis matrix $\mathbf{F}$, there is an important distinction: The basis functions in the time-domain method represent impulse responses, as opposed to general functions in the frequency-domain method in~\cite{JereMILCOM2023}.
Thus, $\mathbf{F}$ computed via the time-domain method can be written as $\mathbf{F} = [\mathbf{f}_0, \mathbf{f}_1, \ldots, \mathbf{f}_m, \ldots, \mathbf{f}_{M-1}]$,
where $\mathbf{f}_m \overset{\mtrian}{=} [f_{0, m} \; f_{1, m} \; \ldots \; f_{N-1, m}]^T \in \mathbb{C}^{N}$ denotes the $m^{\text{th}}$ column of $\mathbf{F}$ and represents a time-domain impulse response. 
Since it is not guaranteed that a specific $\mathbf{f}_{m}$ represents a strictly MP impulse response, we adopt a strategy to transform them into strictly MP impulse responses by adjusting the first tap~\cite{oppenheim2009discrete}.
This is an important step towards constructing stable basis vectors, which is a requirement in the construction of a stable ESN reservoir representing the effective equalizer IIR filter.
Here, we define a `compensation matrix' $\mathbf{B} \in \mathbb{C}^{N \times M}$ as $\mathbf{B} \overset{\mtrian}{=} [\mathbf{b}_0, \; \mathbf{b}_1, \; \ldots, \; \mathbf{b}_{M-1}]$, where $\mathbf{b}_m \overset{\mtrian}{=} [f_{0,m} - b_m, \; 0, \ldots, \; 0]^T = [\mathbf{B}]_{:, m}$ is the $m$-th column of $\mathbf{B}$,
such that the fixed values $\{b_m \}_{m=0}^{M-1}$ are chosen to ensure~\cite{oppenheim2009discrete, Vaidyanathan1993} 
\begin{align}
    b_m > \sum_{n=1}^{N-1} |f_{n,m}|, \; m=0,1,\ldots,M-1.
\end{align} 
The above choice for $\{b_m \}_{m=0}^{M-1}$ ensures that the original basis matrix $\mathbf{F}$ can be decomposed as $\mathbf{F} = \mathbf{P} + \mathbf{B}$, where $\mathbf{P} = [\mathbf{p}_0 ,\; \mathbf{p}_1 ,\; \ldots, \; \mathbf{p}_{M-1}]$, with $\mathbf{p}_m = [b_m, \; f_{1,m}, \; \ldots \; f_{N-1, m}]^T$.
This strategy decomposes potentially mixed-phase impulse responses in $\mathbf{F}$ as a sum of strictly MP impulse responses in $\mathbf{P}$ and a collection of $0$-th order feedforward terms (i.e., `skip' connections corresponding to the $z^0$ term) in $\mathbf{B}$.

\subsection{From Optimum Basis Set to Configured RNN Weights}
\label{sec:optimum_basis_to_weights}
Having obtained the modified strictly MP basis set in $\mathbf{P}$, the next step is to obtain the corresponding weights $\mathbf{W}_{\text{res}}$ and $\mathbf{W}_{\text{in}}$ to configure the reservoir/RNN of the ESN.
Denote $\mathbf{p}_m \in \mathbb{C}^{N}$ as the $m^{\text{th}}$ column of $\mathbf{P}$ and $\widetilde{\mathbf{P}}_m \in \mathbb{C}^{N \times N} = \mathcal{T}(\mathbf{p}_m)$ as its Toeplitz form.
Then, defining $\widetilde{\mathbf{Q}}_m \overset{\mtrian}{=} \widetilde{\mathbf{P}}_{m}^{-1}$, $[\mathbf{Q}_m]_{:,0}$, i.e., the first column of $\widetilde{\mathbf{Q}}_m$ contains the coefficients of the IIR filter whose impulse response is $\mathbf{p}_m$.
In a similar vein, we define $\widetilde{\mathbf{B}}_m \in \mathbb{C}^{N \times N} \overset{\mtrian}{=} \diag{ \{ f_{0,m} - b_m \}_{m=0}^{M-1} }$.

Note that while $[\mathbf{Q}_m]_{:,0} \in \mathbb{C}^{N}$ returns the filter coefficients for an $N$-tap IIR filter, we would ideally like to restrict the order of this IIR filter to only $L_{\mathrm{f}}$ coefficients such that $L_{\mathrm{f}} \ll N$.
This is to ensure that the number of configured weights of the ESN do not scale prohibitively with $M$.
Thus, $L_{\mathrm{f}}$ represents the cut-off point such that the effective `reduced-order' IIR filter is now represented by the truncated filter coefficients $\{q_{n,m} \}_{n=0}^{L_{\mathrm{f}} - 1}$.
The impulse response of the reduced-order filter denoted by $\widehat{\mathbf{p}}_{m}$ approximates the impulse response $\mathbf{p}_{m}$ of the original $N$-tap filter, 
where a reasonable choice of $L_{\mathrm{f}}$ ensures that the approximation error $\|\mathbf{p}_m - \widehat{\mathbf{p}}_m \|_2^2$ is small.
The transfer function of this reduced order IIR filter corresponding to the $m^{th}$ eigenvector is given by
\begin{align}
    Q_m(z) \overset{\mtrian}{=} \frac{1}{\sum_{n=0}^{L_{\mathrm{f}} - 1} q_{n,m}z^{-j}}.
    \label{eq:reduced_order_eigenvector}
\end{align}
Similarly, the transfer function of the skip connection corresponding to the $m^{\text{th}}$ eigenvector can be denoted as $R_m(z) \overset{\mtrian}{=} \frac{1}{f_{0,m} - b_m}$, where
these skip connection weights can be collapsed into a single skip connection in the final ESN implementation.
This results in $\mathbf{w}_{\mathrm{out}} \in \mathbb{C}^{N_{\mathrm{n}} + 1}$ instead of $\mathbf{w}_{\mathrm{out}} \in \mathbb{C}^{N_{\mathrm{n}}}$ with Eq.~\eqref{eq:output_eqn} modified accordingly. 

\begin{figure*}[htbp]
    \centering    \includegraphics[width=0.675\linewidth]{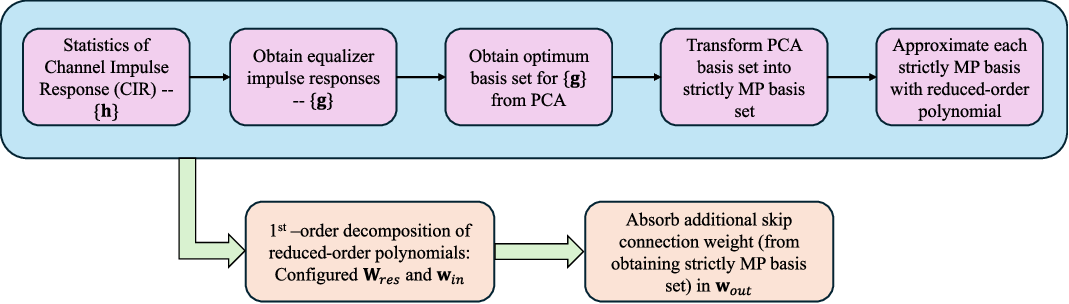}
    \caption{Summary of the time-domain procedure of configuring the untrained RNN weights of the vanilla ESN (``RP'' is shorthand for `Rational Polynomial').}  \label{fig:time_domain_method_flowchart}
\end{figure*}

Since the configured reservoir weights of the ESN correspond to parallel non-interconnected neurons, 
Eq.~\eqref{eq:reduced_order_eigenvector} can be decomposed into a sum of first-order IIR transfer functions as $Q_m(z) \approx \sum_{n=0}^{L_{\mathrm{f}} - 1} \frac{c_{n,m}}{1 - p_{n,m}z^{-1}}$.
Thus, the $m$-th eigenvector is decomposed into $L_{\mathrm{f}}$ first-order poles $\{p_{n,m} \}$ with corresponding weights $\{c_{n,m} \}$. 
For $M$ eigenvectors, we get a total of $ML_{\mathrm{f}}$ neurons in the reservoir with $\mathbf{W}_{\mathrm{res}} = \diag{\{ p_{n,m} \} }$ and $\mathbf{w}_{\mathrm{in}} = \myvec{\{c_{n,m} \}}$ for $m=0,1,\ldots,M-1$ and $n=0,1,\ldots,L_{\mathrm{f}} - 1$.
Thus, the above development completely describes the time-domain approach to obtain the configured weights of the ESN starting with knowledge of the channel statistics.
The time-domain configuration procedure is summarized in the flowchart of  Fig.~\ref{fig:time_domain_method_flowchart}.

\textbf{Block Form of the ESN:}
The configured reservoir/RNN weights of the ESN either via the time-domain method or the frequency-domain now admits a clean ``block form'' interpretation of the ESN with a reservoir consisting of $N_{\mathrm{n}} = K$ parallel recurrent neurons, where $K = M L_{\mathrm{f}}$.
Let $\{p_k \}_{k=0}^{K-1}$ denote the set of configured RNN weights found via either method, then a Vandermonde matrix $\mathbf{\Psi} \in \mathbb{C}^{N \times K}$ can be defined as $\mathbf{\Psi} \overset{\mtrian}{=} [\boldsymbol{\psi}_0 \;, \boldsymbol{\psi}_1 \;, \ldots, \; \boldsymbol{\psi}_{K-1}]$ where $\boldsymbol{\psi}_k \overset{\mtrian}{=} [1, \; p_k, \; \ldots, \; p_{k}^{N-1}]^T = [\mathbf{\Psi}]_{:, k}$.
The $k$-th column of $\mathbf{\Psi}$ represents the time-domain impulse response of the $k$-th first-order IIR filter with pole $p_k$. 
Thus, $\mathbf{\Psi}$ denotes the time-domain dynamics of the constructed reservoir in its atomic form and allows for easier subsequent development.
As an example, assuming linear activation and $\mathbf{w}_{\text{in}} = \boldsymbol{1}$, if $\mathbf{y} \in \mathbb{C}^{N}$ denotes the received time-domain sequence with corresponding Toeplitz form $\widetilde{\mathbf{Y}} \in \mathbb{C}^{N \times N}$ then the reservoir states matrix $\mathbf{S} \in \mathbb{C}^{K \times N}$ can be compactly written in a single step as $\mathbf{S} = (\widetilde{\mathbf{Y}} \mathbf{\Psi})^T$.
Compared to the iterative construction of $\mathbf{S}$ shown in Eq.~\eqref{eq:state_transition_eqn}, the single-step construction makes the analysis more tractable.
This atomic form of the ESN and its associated notation will be adopted in Sec.~\ref{sec:rc_design_objective}.

\section{Configuring Reservoir/RNN Weights for RC-based OFDM Detector}
\label{sec:rc_design_objective}

\subsection{RC Configuration Problem Formulation - Atomic Form}
\label{sec:rc_design_objective_atomic_form}
In this section, we focus on the reservoir/RNN weight configuration problem for RC-based OFDM symbol detector.
Under the symbol detection task, we distinguish between the `configuring' stage and the `training' stage of the ESN.
Specifically, the configuration stage involves computing the untrained input and reservoir/RNN weights of the ESN based on domain knowledge. 
Thus, the configuration stage only uses domain knowledge (channel statistics in our case) to directly set $\mathbf{W}_{\text{res}}$ and $\mathbf{W}_{\text{in}}$, while the training stage involves learning $\mathbf{W}_{\text{out}}$ using only the OTA RSs.
In the following analysis, we assume the ESN to have unit input weights.
In Sec.~\ref{sec:performance_evaluation}, we will show empirically that the insight gained through this analysis still holds even when the ESN employs nonlinear activation.

We proceed to formulate the reservoir/RNN configuration problem as follows.
Denote the transmitted time-domain sequence as $\mathbf{x} \in \mathbb{C}^N$ with its corresponding Toeplitz form denoted by $\widetilde{\mathbf{X}} = \mathcal{T}(\mathbf{x}) \in \mathbb{C}^{N \times N}$.
Let $\mathbf{y}_{\mathrm{o}} \in \mathbb{C}^{N}$ denote the received time-domain sequence over the channel with impulse response $\mathbf{h}_{\mathrm{o}} \in \mathbb{C}^{L}$, so that $\mathbf{y}_{\mathrm{o}} = \widetilde{\mathbf{H}}_{\mathrm{o}} \mathbf{x}$, where $\widetilde{\mathbf{H}}_{\mathrm{o}} = \mathcal{T}([\mathbf{h}_{\mathrm{o}}^T, \boldsymbol{0}_{(N-L) \times 1}^T]^T)$ and $\widetilde{\mathbf{Y}}_{\mathrm{o}} = \mathcal{T}(\mathbf{y}_{\mathrm{o}}) = \widetilde{\mathbf{X}} \widetilde{\mathbf{H}}_{\mathrm{o}}$.
Here, $\widetilde{\mathbf{H}}_{\mathrm{o}}$ also contains the effect of path loss and large-scale shadowing.
In practical receivers, automatic gain control (AGC) is employed to ensure that the average received signal power is maintained at a constant level at the input of the analog-to-digital converter (ADC), thereby effectively compensating for large-scale fading effects such as path loss and shadowing.
This is equivalent to performing the operation $\mathbf{y} = \frac{\mathbf{y}_{\mathrm{o}}}{\|\mathbf{y}_{\mathrm{o}} \|_2} = \frac{\mathbf{y}_{\mathrm{o}}}{\|\widetilde{\mathbf{H}}_{\mathrm{o}} \mathbf{x} \|_2} \approx \frac{\mathbf{y}_{\mathrm{o}}}{\| \mathbf{h}_{\mathrm{o}} \|_2 \| \mathbf{x} \|_2}$, where the approximation is valid for large $N$ and for $L \ll N$~\cite{Zhu2017}.
Thus, the effective input-output relationship is $\mathbf{y} = \widetilde{\mathbf{H}} \mathbf{x}$, where
$\widetilde{\mathbf{H}} = \frac{\widetilde{\mathbf{H}}_{\mathrm{o}}}{\| \widetilde{\mathbf{H}}_{\mathrm{o}} \|_F}$ is the `normalized' Toeplitz channel matrix, resulting in $\|\widetilde{\mathbf{H}} \|_F \approx \sqrt{N} \| \mathbf{h} \|_2 = \sqrt{N}$ when a per-realization normalization $\| \mathbf{h} \|_2 = 1$ is enforced via the AGC. 
Therefore, in the following development, we assume that the channel has been normalized as per above and that the weight configuration procedure utilizes the small-scale fading statistics obtained from such normalized channel realizations.

Ignoring additive noise for the subsequent analysis and denoting the time-domain impulse response of the equalizer as $\mathbf{g}$, the objective is to find the `configured' poles or reservoir/RNN weights $\{p_k \}_{k=0}^{K-1}$ with the corresponding impulse response matrix $\mathbf{\Psi}$ such that the objective function $\mathbb{E}_{\mathbf{h}} [ \| \widehat{\mathbf{x}} - \mathbf{x} \|_2^2]$ is minimized,
where $\widehat{\mathbf{x}}$ is the equalized time-domain sequence.
Without additive noise, the equalized sequence is given by $\widehat{\mathbf{x}} = \widetilde{\mathbf{Y}} \mathbf{g}$.
The corresponding optimization problem becomes
\begin{align}
    \mathbf{g}_{\mathrm{opt}} = \argmin_{\mathbf{g}} \|\widetilde{\mathbf{Y}} \mathbf{g} - \mathbf{x} \|_2^2.
    \label{eq:esn_design_optimization_problem}
\end{align}
Using the impulse response representation admitted by $\mathbf{\Psi}$, we have $\mathbf{g} = \mathbf{\Psi} \mathbf{w}_{\mathrm{out}}$. Substituting this in Eq.~\eqref{eq:esn_design_optimization_problem} and solving for $\mathbf{w}_{\mathrm{out}}$ gives the least-square solution as $\widehat{\mathbf{w}}_{\mathrm{out}} = (\widetilde{\mathbf{Y}} \mathbf{\Psi})^{\dagger} \mathbf{x}$.
Thus, the impulse response $\mathbf{g}$ of the equalizer is given by 
\begin{align}
    \mathbf{g} = \mathbf{\Psi} (\widetilde{\mathbf{Y}} \mathbf{\Psi})^{\dagger} \mathbf{x}.
    \label{eq:equalizer_impulse_response}
\end{align}
Substituting Eq.~\eqref{eq:equalizer_impulse_response} in Eq.~\eqref{eq:esn_design_optimization_problem} and using $\widetilde{\mathbf{Y}} = \widetilde{\mathbf{X}} \widetilde{\mathbf{H}}$, we reach the problem formulation, defined as \textbf{P}, of the reservoir/RNN weight configuration. 
It can be expressed as
\begin{align}
    \textbf{P}:  \min_{\mathbf{\Psi}} \mathbb{E}_{\widetilde{\mathbf{H}}} [\|\widetilde{\mathbf{X}} \widetilde{\mathbf{H}} \mathbf{\Psi} (\widetilde{\mathbf{X}} \widetilde{\mathbf{H}} \mathbf{\Psi})^\dagger \mathbf{x} - \mathbf{x}\|_2^2].
    \label{eq:P1_original}
\end{align}
Problem \textbf{P} in Eq.~\eqref{eq:P1_original} targets to characterize the reservoir/RNN weights for the set of parallel (non-interconnected) recurrent neurons, $\mathbf{\Psi}_{\mathrm{opt}}$, of the ESN.

\subsection{PCA-based Weight Configuration:}
While Problem \textbf{P} attempts to directly identify the parallel recurrent neuron weights $\mathbf{\Psi}_{\mathrm{opt}}$, it is generally more tractable to solve for a basis matrix $\mathbf{F}_{\mathrm{opt}}$ given the domain knowledge (e.g., the statistics) of the target function, which can be decomposed into parallel recurrent neuron weights.
Recall that $\mathbf{F}_{\mathrm{opt}}$ represents the optimum basis matrix for the subspace spanned by the target function under consideration, e.g., the impulse response $\mathbf{g}$ of the time-domain equalizer, as discussed in Sec.~\ref{sec:optimum_basis_pca}.
Furthermore, depending on the nature of the target function and the exact problem formulation, $\mathbf{F}_{\mathrm{opt}}$ can admit multiple solutions.
Therefore, instead of solving Problem \textbf{P} directly, we adopt the strategy of solving the optimization problem in terms of $\mathbf{F}$ and decomposing its solution $\mathbf{F}_{\mathrm{opt}}$, following the procedures outlined in Sec.~\ref{sec:optimum_basis_pca} and Sec.~\ref{sec:optimum_basis_to_weights} to obtain $\mathbf{\Psi}_{\mathrm{opt}}$.

To establish the linkage between Problem \textbf{P} and the optimization problem in terms of $\mathbf{F}$, consider the transmitted vector $\mathbf{x} \in \mathbb{C}^{N}$ to be the unit sample function, i.e., $\mathbf{x} = [1, \, \boldsymbol{0}_{(N-1) \times 1}^T]^T$, leading to $\widetilde{\mathbf{X}} = \mathcal{T}(\mathbf{x}) =  \mathbf{I}_N$. 
Substituting it in the argument of the expectation in Problem \textbf{P}, we have
\begin{align}
    \left\| \left( \widetilde{\mathbf{H}} \mathbf{\Psi} (\widetilde{\mathbf{H}} \mathbf{\Psi})^{\dagger} - \mathbf{I} \right) \mathbf{x} \right\|_2^2 & \overset{(a)}{\leq} \left\| \left( \widetilde{\mathbf{H}}\mathbf{\Psi} (\widetilde{\mathbf{H}} \mathbf{\Psi})^{\dagger} - \mathbf{I} \right)  \right\|_F^2, \\
    &= \left\|  \widetilde{\mathbf{H}} \left( \mathbf{\Psi} (\widetilde{\mathbf{H}} \mathbf{\Psi})^{\dagger} - \mathbf{H}^{-1} \right)  \right\|_F^2,
    \label{eq:P1_modified_Psi}
\end{align}
where $(a)$ follows from the Cauchy-Schwarz inequality and the fact that $\| \mathbf{x} \|_2 = 1$ for this analysis. 
Recall from Sec.~\ref{sec:optimum_basis_pca} and Sec.~\ref{sec:optimum_basis_to_weights} that the basis matrix $\mathbf{F}$ can be decomposed into first-order IIR filter poles, whose impulse responses form the Vandermonde matrix $\mathbf{\Psi}$. 
Thus, any orthogonal basis matrix $\mathbf{F}$ can be expressed as $\mathbf{F} = \mathbf{\Psi} \mathbf{A}$, where $\mathbf{A}$ represents a mixing matrix that constructs every column of $\mathbf{F}$ as a linear combination of the impulse responses of the decomposed poles in $\mathbf{\Psi}$. 
Then, it follows that $\mathcal{F}_{\mathbf{F}} \subseteq \mathcal{F}_{\mathbf{\Psi}}$, where $\mathcal{F}_{\mathbf{F}}$ denotes the linear subspace spanned by the columns of $\mathbf{F}$ and $\mathcal{F}_{\mathbf{\Psi}}$ denotes the linear subspace spanned by the columns of $\mathbf{\Psi}$.
Therefore, the original Problem \textbf{P} given by Eq.~\eqref{eq:P1_original} in terms of $\mathbf{\Psi}$ is first transformed into Eq.~\eqref{eq:P1_modified_Psi}, and can be further related to a problem formulation in terms of $\mathbf{F}$ as follows:
\begin{align*}
    \left\|  \widetilde{\mathbf{H}} \left( \mathbf{\Psi} (\widetilde{\mathbf{H}} \mathbf{\Psi})^{\dagger} - \widetilde{\mathbf{H}}^{-1} \right)  \right\|_F^2 & \overset{(a)}{\leq} \left\|  \widetilde{\mathbf{H}} \left( \mathbf{F} (\widetilde{\mathbf{H}} \mathbf{F})^{\dagger} - \widetilde{\mathbf{H}}^{-1} \right) \right\|_F^2, \nonumber \\
    &= \left\| \widetilde{\mathbf{H}} \left( \mathbf{F} \mathbf{F}^{\dagger}\widetilde{\mathbf{H}}^{-1}  - \widetilde{\mathbf{H}}^{-1} \right) \right\|_F^2, \nonumber \\
    &\overset{(b)}{\leq} \| \widetilde{\mathbf{H}} \|_F^2 \| \mathbf{F} \mathbf{F}^{H}\widetilde{\mathbf{H}}^{-1}  - \widetilde{\mathbf{H}}^{-1}  \|_F^2,
\end{align*}
where $(a)$ follows from $\mathcal{F}_{\mathbf{F}} \subseteq \mathcal{F}_{\mathbf{\Psi}}$ and
$(b)$ follows from the Cauchy-Schwarz inequality and the fact that $\mathbf{F}^{\dagger} = \mathbf{F}^H$ for any orthogonal basis matrix $\mathbf{F}$.
The transformed problem in terms of $\mathbf{F}$, defined as Problem \textbf{P1}, can be expressed as
\begin{align}
    \textbf{P1}: \min_{\mathbf{F}} \mathbb{E}_{\widetilde{\mathbf{H}}} \left[\| \widetilde{\mathbf{H}} \|_F^2 \| \mathbf{F} \mathbf{F}^{H} \widetilde{\mathbf{H}}^{-1}  - \widetilde{\mathbf{H}}^{-1}  \|_F^2 \right],
    \label{eq:F_P1}
\end{align}
with its corresponding solution denoted as $\mathbf{F}^{(\mathrm{P1})}_{\mathrm{opt}}$.
Furthermore, it follows that for ``normalized'' channel realizations adhering to $\| \widetilde{\mathbf{H}} \|_F \approx \sqrt{N}$ as argued in Sec.~\ref{sec:rc_design_objective_atomic_form}, 
minimizing over $\mathbb{E}_{\widetilde{\mathbf{H}}} \left[\| \widetilde{\mathbf{H}} \|_F^2 \| \mathbf{F} \mathbf{F}^{H} \widetilde{\mathbf{H}}^{-1}  - \widetilde{\mathbf{H}}^{-1}  \|_F^2 \right]$ in Problem \textbf{P1} is approximately equivalent to minimizing over $\mathbb{E}_{\widetilde{\mathbf{H}}} \left[ \| \mathbf{F} \mathbf{F}^{H} \widetilde{\mathbf{H}}^{-1}  - \widetilde{\mathbf{H}}^{-1}  \|_F^2 \right]$.
Therefore, Problem \textbf{P1} can be transformed to the equivalent Problem \textbf{P2} defined as
\begin{align}
    \textbf{P2}: \min_{\mathbf{F}} \mathbb{E}_{\widetilde{\mathbf{H}}} \left[ \| \mathbf{F} \mathbf{F}^{H}\widetilde{\mathbf{H}}^{-1}  - \widetilde{\mathbf{H}}^{-1}    \|_F^2 \right],
    \label{eq:F_P2}
\end{align}
with its corresponding solution denoted as $\mathbf{F}^{(\mathrm{P2})}_{\mathrm{opt}}$.
Recognizing that $\widetilde{\mathbf{H}}^{-1} = \mathcal{T}(\mathbf{g})$, Problem \textbf{P2} can be equivalently written as 
\begin{align}
    \textbf{P2}: \min_{\mathbf{F}} \mathbb{E}_{\mathbf{g}} \left[ \| \mathbf{F} \mathbf{F}^{H} \mathcal{T}(\mathbf{g})  - \mathcal{T}(\mathbf{g})    \|_F^2 \right].
    \label{eq:F_P2_Toeplitz}
\end{align}

Accordingly, the original Problem $\textbf{P}$ in terms of $\mathbf{\Psi}$ has been transformed to the formulation of Problem $\textbf{P2}$ in terms of $\mathbf{F}$. 
We can further link Problem \textbf{P2} to the standard dimensionality reduction formulation by setting the target function as $\mathbf{g}$.
In this way, the problem of finding the optimum basis matrix for the linear subspace spanned by $\mathbf{g}$, defined as Problem
\textbf{P*}, can be expressed as
\begin{align}
    \textbf{P*}: \min_{\mathbf{F}} \mathbb{E}_{\mathbf{g}} [\|\mathbf{F} \mathbf{F}^H \mathbf{g} - \mathbf{g} \|_2^2].
\label{eq:pca_optimization_time_domain}
\end{align}

Note that the solution of Problem \textbf{P*} can be obtained through PCA denoted as $\mathbf{F}_{\mathrm{opt}}^{(\mathrm{P^{*}})}$. 
It is orthogonal satisfying $\left( \mathbf{F}_{\mathrm{opt}}^{(\mathrm{P^{*}})} \right)^H  \mathbf{F}_{\mathrm{opt}}^{(\mathrm{P^{*}})} = \mathbf{I}_M$.
Through this process we have achieved the objective of linking Problem \textbf{P} in terms of $\mathbf{\Psi}$ to Problem \textbf{P*} in terms of $\mathbf{F}$ via the chain \textbf{P} $\rightarrow$ \textbf{P1} $\rightarrow$ \textbf{P2} $\rightarrow$ \textbf{P*}.
In this way, we will be able to solve Problem \textbf{P*} which has a well-established solution $\mathbf{F}_{\mathrm{opt}}^{(\mathrm{P}^{*})}$ given by PCA, and utilize the solution to configure the reservoir/RNN weights of the ESN following the procedure outlined in Sec.~\ref{sec:optimum_basis_pca} and Sec.~\ref{sec:optimum_basis_to_weights}.

Furthermore, we are interested in analytically characterizing the approximation error of the ESN achieved by utilizing $\mathbf{F}_{\mathrm{opt}}^{(\mathrm{P}^{*})}$ in the objective function of Problem $\textbf{P2}$.
To this end, we first introduce the following lemma.

\begin{lemma}
\label{lemma:pca_min_error}
The minimum approximation error achieved by $\mathbf{F}_{\mathrm{opt}}^{(\mathrm{P}^{*})}$ in the objective of Problem \textbf{P*}~\eqref{eq:pca_optimization_time_domain} is
\begin{align}
    \mathbb{E}_{\mathbf{g}} \left[ \left\| \mathbf{F}_{\mathrm{opt}}^{(\mathrm{P}^{*})} \left(\mathbf{F}_{\mathrm{opt}}^{(\mathrm{P}^{*})} \right)^{H} \mathbf{g} - \mathbf{g} \right\|_2^2 \right] = \sum_{j=M}^{N-1} \lambda_j,
\end{align}
where $\lambda_j$ is the $j$-th eigenvalue of $\mathbf{K} = \mathbb{E}[\mathbf{g} \mathbf{g}^H]$.
\end{lemma}
\textit{Proof}. 
See Appendix~\ref{sec:Appendix_A}.

Accordingly, the approximation error achieved by utilizing $\mathbf{F}_{\mathrm{opt}}^{(\mathrm{P}^{*})}$ to solve Problem \textbf{P2} can be characterized in Theorem~\ref{theorem:bound_on_error}. 
\begin{thm} \label{theorem:bound_on_error}
The approximation error $\varepsilon_{\mathrm{app}}^{(\mathrm{P2})} $ achieved by $\mathbf{F}_{\mathrm{opt}}^{(\mathrm{P}^{*})}$ in the objective of Problem $\textbf{P2}$ in Eq.~\eqref{eq:F_P2_Toeplitz} is
\begin{align*}       
    \varepsilon_{\mathrm{app}}^{(\mathrm{P2})} = \sum_{i=0}^{N-1} \left[ \Tr(\mathbf{K} \boldsymbol{L}_i^H \boldsymbol{L}_i ) - \Tr\left( \mathbf{K} \boldsymbol{L}_i^H \mathbf{F}_{\mathrm{opt}}^{(\mathrm{P}^{*})} (\mathbf{F}_{\mathrm{opt}}^{(\mathrm{P}^{*})})^{H} \boldsymbol{L}_i \right) \right],
%    \label{eq:thm_equation}
\end{align*}
where $\mathbf{K} = \mathbb{E}[\mathbf{g} \mathbf{g}^H]$ and  $\boldsymbol{L}_i \overset{\mtrian}{=}{[\boldsymbol{0}, \boldsymbol{0}; \mathbf{I}_{N-i}, \boldsymbol{0}]} \in \mathbb{R}^{N \times N}$ denotes the shift matrix.
\end{thm} 
\textit{Proof}. 
See Appendix~\ref{sec:Appendix_B}.

\begin{figure}[htbp]
    \centering    \includegraphics[width=0.75\linewidth]{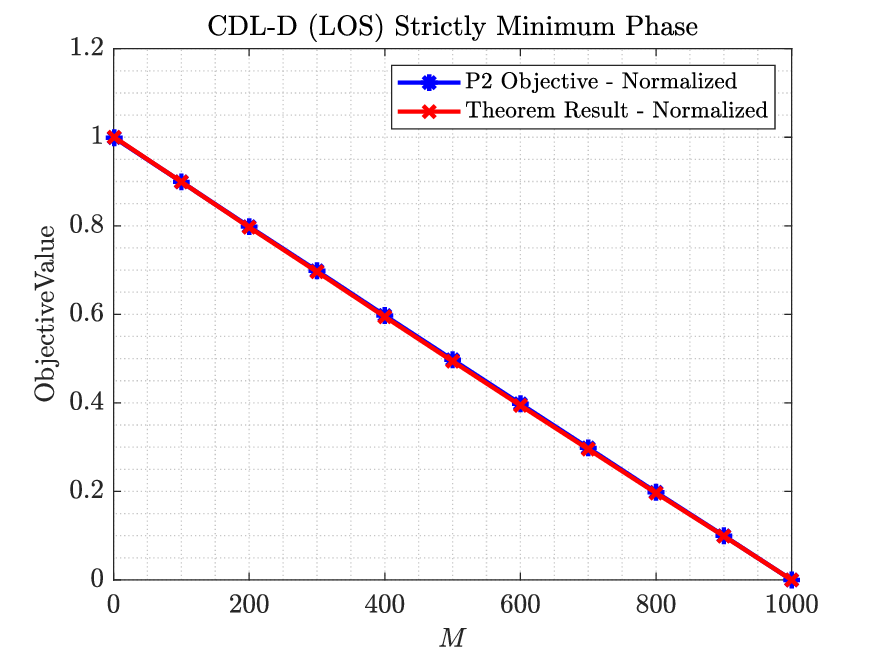}
    \caption{Validation of Theorem~\ref{theorem:bound_on_error} via numerical evaluation.}  \label{fig:Theorem_validation}
\end{figure}

From the problem chain \textbf{P} $\rightarrow$ \textbf{P1} $\rightarrow$ \textbf{P2} $\rightarrow$ \textbf{P*} we know that the approximation error characterized in Theorem~\ref{theorem:bound_on_error} is an upper-bound of the minimum achievable approximation error of Problem \textbf{P}. 
Therefore, Theorem~\ref{theorem:bound_on_error} provides theoretical performance guarantees on the approximation error of utilizing the introduced reservoir/RNN weight configuration procedure for the ESN.
Meanwhile, Fig.~\ref{fig:Theorem_validation} validates Theorem~\ref{theorem:bound_on_error} through numerical evaluation for sequence length $N=1000$ with $N_{\mathrm{obs}} = 1000$ strictly MP realizations of the 3GPP clustered delay
line-D (CDL-D) channel. 
Denoting $\widetilde{\mathbf{G}}_{(r)} = (\widetilde{\mathbf{H}}^{-1})_{(r)}$ as the Toeplitz form of the equalizer impulse response for the $r$-th channel realization, we plot $\frac{ \frac{1}{N_{\mathrm{obs}}} \sum_{r=1}^{N_{\mathrm{obs}}} \| \mathbf{F} \mathbf{F}^{H}\widetilde{\mathbf{G}}_{(r)}  - \widetilde{\mathbf{G}}_{(r)}    \|_F^2}{ \frac{1}{N_{\mathrm{obs}}} \sum_{r=1}^{N_{\mathrm{obs}}} \| \widetilde{\mathbf{G}}_{(r)} \|_F^2}$ and compare it against $\frac{\sum_{i=0}^{N-1} [\Tr(\widehat{\mathbf{K}} \boldsymbol{L}_i^H \boldsymbol{L}_i ) - \Tr(\widehat{\mathbf{K}} \boldsymbol{L}_i^H \mathbf{F}\mathbf{F}^H \boldsymbol{L}_i)]}{\frac{1}{N_{\mathrm{obs}}} \sum_{r=1}^{N_{\mathrm{obs}}} \| \widetilde{\mathbf{G}}_{(r)} \|_F^2}$, where $\widehat{\mathbf{K}} = \frac{1}{N_{\mathrm{obs}}} \sum_{r=1}^{N_{\mathrm{obs}}} \mathbf{g}_{(r)} \mathbf{g}_{(r)}^H$ is the empirical covariance matrix.
We can observe from Fig.~\ref{fig:Theorem_validation} that the normalized numerically evaluated value based on Eq.~\eqref{eq:F_P2} matches well with the normalized theoretical expression characterized in Theorem~\ref{theorem:bound_on_error} across $M$, the number of most significant eigenvectors used in PCA.
Furthermore, both normalized values monotonically decrease with $M$ while satisfying $\varepsilon_{\mathrm{app}}^{(\mathrm{P2})} = 0$ for $M=N$, as expected.
Since $M$ and the reduced filter order $L_{\mathrm{f}}$ determine the number of reservoir/RNN neurons in both frequency/time-domain weight configuration methods, Theorem~\ref{theorem:bound_on_error} shows a clear trade-off between the achieved approximation error and the number of neurons in the reservoir/RNN, thus providing operational guidance on setting the size of reservoir/RNN for RC-based approaches.

\section{Model Explainability for WESN}
\label{sec:nmp_channels_wesn_connection}
\subsection{Approximate Stable Inverse for Mixed-phase Systems}

A general $L$-tap mixed-phase channel with transfer function $H(z)$ comprises both strictly MP and strictly NMP factors, i.e., it can be factorized as $H(z) = H_{\mathrm{MP}}(z) H_{\mathrm{NMP}}(z)$, where $H_{\mathrm{MP}}(z)$ denotes the strictly MP factor of $H(z)$ and $H_{\mathrm{NMP}}(z)$ denotes its strictly NMP factor, implying that the roots of $H_{\mathrm{MP}}(z)$ lie strictly inside the unit circle and those of $H_{\mathrm{NMP}}(z)$ lie strictly outside the unit circle. 
Then, a \emph{stable} inverse for this mixed-phase channel is approximated in the transform domain as~\cite{Moir2022}: $H^{-1}(z) \approx \sum_{\ell=0}^{L-1} \frac{c_\ell}{1 - p_{\ell}z^{-1}} + \sum_{i=0}^{L_{ff}-1} k_i z^{-i}$,
where the feedforward terms $\{k_i \}_{i=0}^{L_{ff}-1}$ represent an FIR filter component of order $L_{ff}$. 
This results in $H^{-1}(z)H(z) \approx z^{-(L+L_{ff}-1)}$, implying that the equalized output incurs a delay of $(L+L_{ff}-1)$ samples. 
To summarize, a strictly MP channel (FIR filter) can be perfectly equalized by an all-pole IIR filter if the channel is perfectly known.
Such an all-pole IIR filter can be decomposed as a sum of first-order IIR filters. 
Therefore, perfect knowledge of the channel FIR filter results in a perfect equalizer that suffices to only use the IIR filter forms.
However, the channel impulse response (CIR) cannot be perfectly known in practice.
In such a scenario, the equalizer performance can be greatly improved by including feedforward taps whose weights are trained using the available training data~\cite{Moir2022}.
This is also confirmed from the fact that the transfer function of an all-pole IIR filter can be approximated as an expansion of feedforward terms of the form $z^{-i}$ via long division~\cite{Moir2022, Vaidyanathan1993}.
Furthermore, for strictly NMP or mixed-phase channels, a \emph{stable} equalizer must include both an IIR filter component as well as an FIR filter component for good equalization performance with modest filter orders~\cite{Moir2022}, even with perfect channel knowledge.

\subsection{Explainability and Weight Configuration for WESN}
\label{sec:connection_with_wesn}

As noted earlier, the stable inverse filter that equalizes a mixed-phase channel generally consists of both an IIR filter component (all-pole form) as well as an FIR filter component (all-zero form)~\cite{Moir2022}.
This foundational fact can be immediately tied to the introduced WESN architecture~\cite{zhou2019}, where it was observed that the WESN significantly outperforms the vanilla ESN in symbol detection performance using randomly generated weights $\mathbf{W}_{\text{in}}$ and $\mathbf{W}_{\text{res}}$.
For explainability, the WESN architecture can be divided into two major components: i) Input windowing, and ii) Output skip and delayed connections.
Input windowing, which was shown to improve the short-term memory of the WESN~\cite{zhou2019}, also has an implicit FIR filtering effect via the moving input window.
On the other hand, the output skip and delayed connections have an explicit FIR filtering effect, and thus have greater impacts in improving equalization performance and subsequent symbol detection metrics such as bit error rate (BER).
Therefore, the symbol detection performance improvement of the WESN over the vanilla ESN is primarily due to output skip and delayed connections.
For the WESN architecture with the output skip and delayed connections, the output weights matrix changes as $\mathbf{W}_{\mathrm{out}} \in \mathbb{C}^{d_{\mathrm{out}} \times (N_{\mathrm{n}} + d_{\mathrm{in}} N_{w})}$, where $N_w$ denotes the `window' length.
Consequently, the output equation for the WESN changes according to $\mathbf{x}_{\mathrm{out}}[n] = \mathbf{W}_{\mathrm{out}} [\mathbf{x}_{\mathrm{res}}[n]^T, \mathbf{x}_{\mathrm{in}}[n]^T, \mathbf{x}_{\mathrm{in}}[n-1]^T, \ldots, \mathbf{x}_{\mathrm{in}}[n-N_w + 1]^T ]^T$.
Note that the elements in $\mathbf{W}_{\mathrm{out}}$ corresponding to the $N_{\mathrm{n}}$ reservoir/RNN neurons linearly combine the reservoir's IIR filter output, while the elements in $\mathbf{W}_{\mathrm{out}}$ corresponding to $N_w$ linearly combine the current and delayed versions of the input directly, thereby applying a weighted tap delay line given by $\{z^0, z^{-1}, \ldots, z^{-(N_w - 1)} \}$ to the input and performing an \emph{explicit} FIR filtering operation.
Finally, the WESN architecture of~\cite{zhou2019} also implements `delay learning' using the training data to iteratively find the optimum delay $D$ that minimizes $\oint_{|z|=1} | H^{-1}(z) H(z) - z^{-D} |^2 dz$, thus aligning well with the delay incurred in the stable inverse of a mixed-phase wireless channel~\cite{Moir2022}, for which $d_{\mathrm{in}} = d_{\mathrm{out}} = 1$.
Thus, both attributes of the WESN~\cite{zhou2019} architecture namely, output skip and delayed connections and delay learning, closely match signal processing fundamentals.
In Sec.~\ref{sec:performance_evaluation}, we will consider the WESN architecture with the output skip and delay connections instead of the vanilla ESN discussed in Sec.~\ref{sec:conventional_esn}.

The reservoir/RNN weight configuration procedure of WESN for a general mixed-phase channel can be described as the following.
Recall that a general mixed-phase channel can be factorized as $H(z) = H_{\mathrm{MP}}(z) H_{\mathrm{NMP}}(z)$.
Since the strictly MP factor $H_{\mathrm{MP}}(z)$ can be equalized by an all-pole IIR filter $\frac{1}{H_{\mathrm{MP}}(z)}$ and due to this direct inverse equalizer being stable, the untrained weights $\mathbf{W}_{\mathrm{in}}$ and $\mathbf{W}_{\mathrm{res}}$ can be configured using the statistics of $H_{\mathrm{MP}}(z)$ alone, using either the frequency-domain method of~\cite{JereMILCOM2023} or the time-domain method introduced in Sec.~\ref{sec:optimum_basis_pca} and Sec.~\ref{sec:optimum_basis_to_weights}.
The output weights corresponding to the additional skip and delayed feedforward taps in the WESN can be trained using the training data (e.g., OTA RSs) in each slot to equalize the strictly NMP factor $H_{\mathrm{NMP}}(z)$.
In this manner, the equalization capability of the WESN architecture can be attributed to each of its constituent components. 
This clearly explains why, for a general mixed-phase wireless channels (e.g., non-line-of-sight channels), WESN-based approaches significantly outperforms vanilla ESN-based approaches.
This clearly attributable explainability as well as the ability to configure (rather than train) certain components of an NN-based architecture using domain knowledge is especially important for NextG PHY receive processing.

\section{Configuring Weights for RC-based MIMO-OFDM Symbol Detectors}
\label{sec:mimo_equalization}

In this section, we extend the reservoir/RNN weight configuration analysis to MIMO-OFDM systems. 
Consider a MIMO system with $N_{\mathrm{t}}$ transmit antennas and $N_{\mathrm{r}}$ receive antennas.
For a frequency-selective MIMO channel with $L$ delay taps, the MIMO coefficient matrix for the $\ell$-th delay tap is denoted as $\boldsymbol{H}_{\ell} \in \mathbb{C}^{N_{\mathrm{r}} \times N_{\mathrm{t}}}$.
The corresponding transfer function for the MIMO channel is given by $\mathbf{H}(z) = \sum_{\ell=0}^{L-1} \boldsymbol{H}_{\ell} z^{-\ell}$.
Thus, for the MIMO channel which is represented as a MIMO FIR filter, the WESN architecture provides a systematic way of incorporating both a MIMO IIR component as well as a MIMO FIR component, which are both instrumental in its equalization process of the MIMO channel~\cite{Vaidyanathan1993} to achieve good MIMO-OFDM symbol detection performance.
To develop a weight configuration procedure for the WESN in the MIMO setting, we start with the case study of a simplistic MIMO channel in Sec.~\ref{sec:factorizable_mimo_channel} to lay the foundation of the overall configuration procedure.
General MIMO channels are considered in Sec.~~\ref{sec:mimo_parametric_channel_model} using the parametric channel representation.

\subsection{Factorizable MIMO Channel: A Simple Case Study}
\label{sec:factorizable_mimo_channel}

The simplistic MIMO channel considered in this case study can be written as $\mathbf{H}(z) = \boldsymbol{H}_{0} h(z)$,
where $h(z) = h_0 + h_1 z^{-1} + \ldots + h_{L-1}z^{-(L-1)}$ is a single input single output (SISO) channel impulse response and hence its description as being `factorizable'.
We assume that $h(z)$ is strictly MP in the following development.  
The same analysis also holds by replacing $h(z)$ with its strictly MP factor $h_{\mathrm{MP}}(z)$ if $h(z)$ is mixed-phase.
Continuing with the strictly MP assumption, the transfer function of the ideal inverse of the channel is given by $\mathbf{G}(z) = \boldsymbol{H}_0^{-1} \frac{1}{h(z)}$.
To better elucidate the details of the weight configuration procedure, we consider the example of a $2 \times 2$ MIMO channel, i.e., $N_{\mathrm{t}} = N_{\mathrm{r}} = 2$.
Let $\boldsymbol{H}_0$ be defined as $\boldsymbol{H}_0 \overset{\mtrian}{=} [h_{11},  h_{12}; h_{21},  h_{22}]$.
Assuming that $\boldsymbol{H}_0$ is invertible, define its inverse as $\boldsymbol{H}_0^{-1} \overset{\mtrian}{=} [g_{11}, g_{12} ; g_{21},  g_{22}]$.
Then, for the transmitted symbol vector $\mathbf{x}[n] \in \mathbb{C}^{N_{\mathrm{t}}}$, the received symbol vector $\mathbf{y}[n] \in \mathbb{C}^{N_{\mathrm{r}}}$, ignoring AWGN, is given by $\mathbf{y}[n] \overset{\mtrian}{=} [y_1[n]; y_2[n]] = \sum_{\ell=0}^{L-1} \boldsymbol{H}_{\ell} \mathbf{x}[n- \ell]$.

\begin{figure}[h]
    \centering
    \subfloat[Step 1]{
        \includegraphics[width=0.4\textwidth]{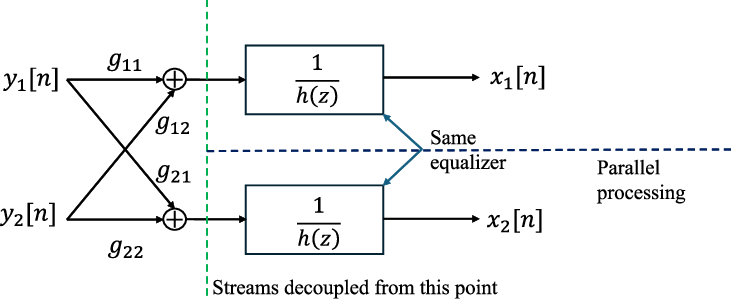}
        \label{fig:factorizable_mimo_fig1}
    }\\ % Line break for vertical arrangement
    \subfloat[Step 2]{
        \includegraphics[width=0.325\textwidth]{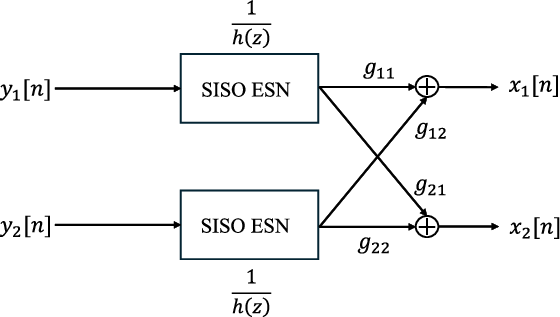}
        \label{fig:factorizable_mimo_fig3}
    }
    \caption{ESN configuration for ``factorizable'' MIMO channel: Steps 1 and 2.}
\end{figure}

The equalizer architecture that perfectly equalizes $\mathbf{y}[n]$ is depicted in Fig.~\ref{fig:factorizable_mimo_fig1}.
Here, we make the following important observations.
First, the weights in $\boldsymbol{H}_0^{-1}$ to invert the effect of $\boldsymbol{H}_0$ are in the front of the processing chain.
Second, the two streams are completely decoupled after acted upon by $\boldsymbol{H}_0^{-1}$.
Finally, due to the factorizable nature of the MIMO channel considered, the SISO equalizers for each stream are the same, i.e., $\frac{1}{h(z)}$.
Next, due to linearity of the operations involved and since the equalizer in each path is the same, the equalizer structure can be modified as shown in Fig.~\ref{fig:factorizable_mimo_fig3}.
The main motivation to perform this transformation is to move the channel-specific equalizer weights in $\boldsymbol{H}_0^{-1}$ closest to the output, thereby aligning with the architecture of the vanilla ESN and the WESN where the output weights are trained with online training data (e.g., OTA RSs).

\begin{figure*}[htbp]
    \centering
    
    \subfloat[Parallel processing using two separate SISO ESNs]{\includegraphics[width=0.3825\linewidth, height=0.1215\linewidth]{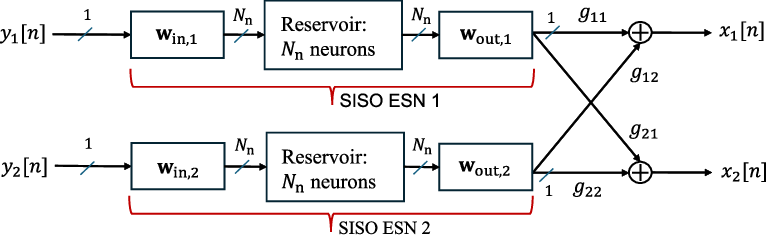}\label{fig:factorizable_mimo_fig4}}%
    \qquad
    \subfloat[A single composite MIMO ESN]{\includegraphics[width=0.3825\linewidth, height=0.1215\linewidth]{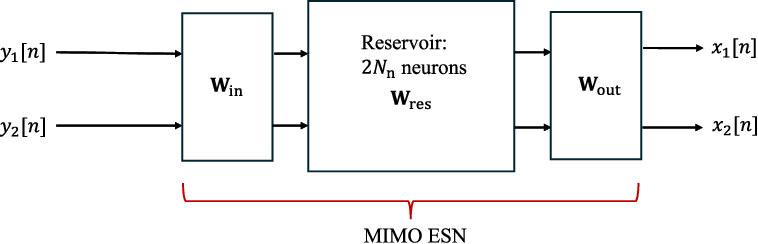}\label{fig:factorizable_mimo_fig5}}%
    \caption{ESN configuration for a ``factorizable'' MIMO channel: Steps 3 and 4.}
\end{figure*}

With this transformed architecture of the $2 \times 2$ MIMO equalizer, each SISO equalizer $\frac{1}{h(z)}$ can be implemented as a `SISO ESN'.
The untrained weights $\mathbf{W}_{\mathrm{res}}$ and $\mathbf{w}_{\mathrm{in}}$ of each SISO ESN can be configured using the empirical statistics of $h(z)$ via the frequency-domain or time-domain methods.
Denoting the respective weights for the $i^{\mathrm{th}}$ SISO WESN with $N_{\mathrm{n}}$ reservoir/RNN neurons as $\mathbf{W}_{\mathrm{res}, i} \in \mathbb{C}^{N_{\mathrm{n}} \times N_{\mathrm{n}}}$, $\mathbf{w}_{\mathrm{in}, i} \in \mathbb{C}^{N_{\mathrm{n}} \times 1}$, $\mathbf{w}_{\mathrm{out}, i} \in \mathbb{C}^{1 \times N_{\mathrm{n}}}$ and the state vector as $\mathbf{s}_{i}[n] \in \mathbb{C}^{N_{\mathrm{n}}}$, the combined state update equation for the two SISO ESNs in parallel can be written as
$[\mathbf{s}_1[n]; \mathbf{s}_2[n]] = \blkdiag(\mathbf{W}_{\mathrm{res}, 1}, \mathbf{W}_{\mathrm{res}, 2}) [[\mathbf{s}_1[n-1]; \mathbf{s}_2[n-1]]] + \blkdiag(\mathbf{w}_{\mathrm{in}, 1}, \mathbf{w}_{\mathrm{in}, 2}) [y_1[n]; y_2[n]]$.
The associated output equation can then be written as $[\widehat{x}_1[n]; \widehat{x}_2[n]] = \boldsymbol{H}_0^{-1} \blkdiag(\mathbf{w}_{\mathrm{out},1}, \mathbf{w}_{\mathrm{out},2}) [\mathbf{s}_1[n]; \mathbf{s}_2[n]]$.
Thus, the `effective' output matrix $\mathbf{W}_{\mathrm{out,eff}} \in \mathbb{C}^{2 \times 2 N_{\mathrm{n}}}$ 
% which does not necessarily exhibit a particular structure, 
becomes $\mathbf{W}_{\mathrm{out,eff}} = [g_{11} \mathbf{w}_{\mathrm{out}, 1}, \, g_{12} \mathbf{w}_{\mathrm{out}, 2}; \, g_{21} \mathbf{w}_{\mathrm{out}, 1}, \, g_{22} \mathbf{w}_{\mathrm{out}, 2}]$.
Since the two SISO ESNs equalize the same channel $h(z)$ with a single unique PDP and thereby single unique channel statistics, the configured weights obey $\mathbf{W}_{\mathrm{res}, 1} = \mathbf{W}_{\text{res}, 2}$ and $\mathbf{w}_{\mathrm{in}, 1} = \mathbf{w}_{\mathrm{in}, 2}$.
The two SISO ESNs with $N_{\mathrm{n}}$ reservoir/RNN neurons each can be collapsed into a single MIMO ESN with $2 N_{\mathrm{n}}$ reservoir/RNN neurons with its configured weights given by $\mathbf{W}_{\mathrm{res, eff}} = \blkdiag(\mathbf{W}_{\mathrm{res}, 1}, \mathbf{W}_{\mathrm{res}, 1})$ and $\mathbf{W}_{\mathrm{in, eff}} = \blkdiag(\mathbf{w}_{\mathrm{in}, 1}, \mathbf{w}_{\mathrm{in}, 1})$.
These steps are illustrated in Fig.~\ref{fig:factorizable_mimo_fig4} and Fig.~\ref{fig:factorizable_mimo_fig5}. 
Finally, input windowing and output skip and delayed connections with window length $N_w$ are added to obtain the configured WESN for the MIMO system.

\subsection{Symbol Detection for General MIMO-OFDM}
\label{sec:mimo_parametric_channel_model}
In this section, we consider general MIMO systems based on the parametric channel representation.
Specifically, we consider the point-to-point MIMO communication scenario with uniform linear arrays (ULAs) deployed at both the transmitter and the receiver.
For a ULA with $N_{\mathrm{t}}$ antenna elements, its array steering vector is given by $\mathbf{a}(\theta) \overset{\mtrian}{=} \left[ 1, \; e^{j2\pi \frac{d \cos(\theta)}{\lambda_c}}, \; \ldots \; ,e^{j2\pi (N_{\mathrm{t}} - 1)\frac{d \cos(\theta)}{\lambda_c}} \right]^T$,
where $d$ is the spacing between antenna elements, $\lambda_c$ is the carrier wavelength and $\theta$ denotes the azimuth angle of arrival (AoA) or angle of departure (AoD), measured relative to the axis of the ULA. 
For a frequency-selective channel with $L$ delay taps, the MIMO coefficient matrix for the $\ell^{\text{th}}$ tap under the parametric MIMO channel representation can be written as $\boldsymbol{H}_{\ell} = \frac{1}{\sqrt{N_{\mathrm{path}}^{(\ell)}}} \sum_{q=0}^{N_{\mathrm{path}}^{(\ell)} - 1} c_{q}^{(\ell)} \mathbf{a}_{\mathrm{r}}(\theta_{\mathrm{r}, q}) \mathbf{a}_{\mathrm{t}}^{T}({\theta_{\mathrm{t}, q}})$,
where $\mathbf{a}_{\mathrm{r}}(\cdot)$ and $\mathbf{a}_{\mathrm{t}}(\cdot)$ are the array steering vectors for the receive and transmit ULAs. 
Here, $N_{\mathrm{path}}^{(\ell)}$ denotes the number of propagation paths through which signals corresponding to the $\ell^{\text{th}}$ delay tap travel from the transmitter antenna array to the receiver antenna array.
$\theta_{\mathrm{r}, q}$ and $\theta_{\mathrm{t}, q}$ denote the AoA and AoD respectively for the $q^{\text{th}}$ path.
$c_{q}^{(\ell)}$ denotes the complex channel gain for the $q^{\text{th}}$ path that falls onto the $\ell^{\text{th}}$ delay tap. 
The MIMO channel transfer function can be written as $\mathbf{H}(z) = \sum_{\ell=0}^{L-1} \boldsymbol{H}_{\ell} z^{-\ell}$, 
where $\rank(\boldsymbol{H}_{\ell}) \leq \min (N_\mathrm{t}, N_\mathrm{r}, N_\mathrm{path}^{(\ell)})$.
In this work, we assume $N_{\mathrm{path}}^{(\ell)} \geq \min(N_{\mathrm{t}}, N_{\mathrm{r}})$, so that $\boldsymbol{H}_{\ell}$ is full-rank with probability $1$. 
This condition is generally satisfied in the FR1 band (sub-6 GHz), whereas in the FR2 (mmWave) and FR3 bands, we may have $N_{\mathrm{path}}^{(\ell)} < \min(N_{\mathrm{t}}, N_{\mathrm{r}})$, especially with massive MIMO systems.

For subsequent analysis, we consider the simplification that $N_{\mathrm{path}}^{(\ell)} = N_{\mathrm{path}}$ for $\ell=0,1,\ldots,L-1$.
This adheres to the modeling of MIMO channels in 3GPP standards, e.g., Spatial Channel Model (SCM)~\cite{3gppTR25996}, for which $N_{\mathrm{path}} = 20$ and $L=6$.
Using the shorthand notation $N_{\mathrm{p}} \overset{\mtrian}{=} N_{\mathrm{path}}$, $\mathbf{a}_{\mathrm{r}}(\theta_{\mathrm{r}, q}) \overset{\mtrian}{=} \mathbf{a}_{\mathrm{r}, q}$ and $\mathbf{a}_{\mathrm{t}}(\theta_{\mathrm{t}, q}) \overset{\mtrian}{=} \mathbf{a}_{\mathrm{t}, q}$, $\mathbf{H}(z)$ can be expanded as~\cite{RubayetDoA}
\begin{align*}
    &\mathbf{H}(z) = \nonumber \\
    & c_{0}^{(0)} \mathbf{a}_{\mathrm{r}, 0} \mathbf{a}_{\mathrm{t}, 0}^T + \left(c_{0}^{(1)} \mathbf{a}_{\mathrm{r}, 0} \mathbf{a}_{\mathrm{t}, 0}^T \right) z^{-1} + \ldots + 
    \left(c_{0}^{(L)} \mathbf{a}_{\mathrm{r}, 0} \mathbf{a}_{\mathrm{t}, 0}^T \right) z^{-L} \nonumber \\
    &+ c_{1}^{(0)} \mathbf{a}_{\mathrm{r}, 1} \mathbf{a}_{\mathrm{t}, 1}^T + \left(c_{1}^{(1)} \mathbf{a}_{\mathrm{r}, 1} \mathbf{a}_{\mathrm{t}, 1}^T \right) z^{-1} + \ldots + 
    \left(c_{1}^{(L)} \mathbf{a}_{\mathrm{r}, 1} \mathbf{a}_{\mathrm{t}, 1}^T \right) z^{-L} \nonumber \\
    &+ \ldots + \nonumber \\
    & c_{N_{\mathrm{p}} - 1}^{(0)} \mathbf{a}_{\mathrm{r}, N_{\mathrm{p}} - 1} \mathbf{a}_{\mathrm{t}, N_{\mathrm{p}} - 1}^T + \ldots + 
    \left(c_{N_\mathrm{p} - 1}^{(L)} \mathbf{a}_{\mathrm{r}, N_\mathrm{p} - 1} \mathbf{a}_{\mathrm{t}, N_\mathrm{p} - 1}^T \right) z^{-L}.    
\end{align*}
Defining $\boldsymbol{A}_{\mathrm{r}} \overset{\mtrian}{=} [\mathbf{a}_{\mathrm{r}, 0}, \, 
\mathbf{a}_{\mathrm{r}, 1},  \ldots,
\mathbf{a}_{\mathrm{r}, N_{\mathrm{p}} - 1}] \in \mathbb{C}^{N_{\mathrm{r}} \times N_{\mathrm{p}}}$, 
$\boldsymbol{A}_{\mathrm{t}} \overset{\mtrian}{=} [\mathbf{a}_{\mathrm{t}, 0}, \, 
\mathbf{a}_{\mathrm{t}, 1}, \, \ldots,
\mathbf{a}_{\mathrm{t}, N_{\mathrm{p}} - 1}] \in \mathbb{C}^{N_{\mathrm{t}} \times N_{\mathrm{p}}}$ and
$\mathbf{D}(z) = \diag \left( \sum_{\ell=0}^{L} c_{0}^{(\ell)} z^{-\ell}, \, \sum_{\ell=0}^{L} c_{1}^{(\ell)} z^{-\ell}, \, \ldots, \, 
\sum_{\ell=0}^{L} c_{N_{\mathrm{p}} - 1}^{(\ell)} z^{-\ell} \right)$ to be an $N_{\mathrm{p}} \times N_{\mathrm{p}}$ transfer function matrix, $\mathbf{H}(z)$ can be compactly written as $\mathbf{H}(z) = \boldsymbol{A}_{\mathrm{r}} \mathbf{D}(z) \boldsymbol{A}_{\mathrm{t}}^T$~\cite{RubayetDoA}.
This compact representation is useful to develop the weight configuration strategy for the WESN.

\begin{figure}[h]
    \centering    \includegraphics[width=\linewidth]{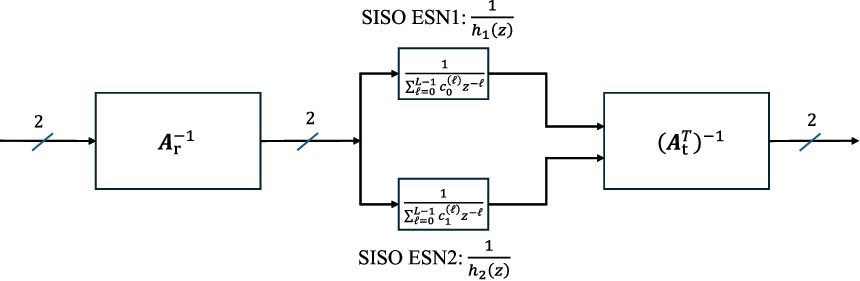}
    \caption{ESN configuration for parametric $2 \times 2$ MIMO channel: Step 1.}
    \label{fig:parametric_mimo_2x2_fig1}
\end{figure}

\begin{figure*}[htbp]
    \centering
    
    \subfloat[Parallel processing using two separate SISO ESNs]{\includegraphics[width=0.42\linewidth, height=0.155\linewidth]{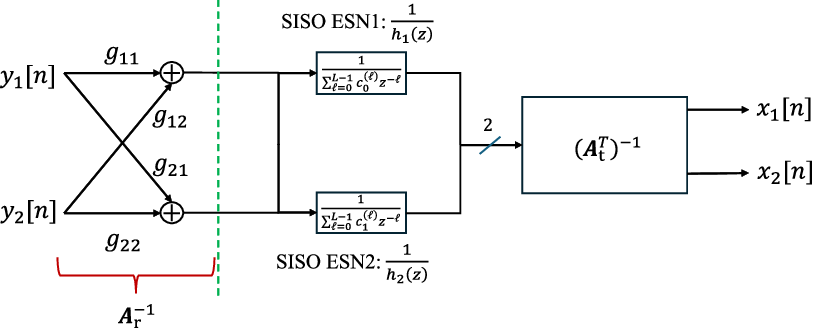}\label{fig:parametric_mimo_2x2_fig2}}%
    \qquad
    \subfloat[Replication of SISO ESNs due to distinct SISO channel transfer functions in the parametric MIMO channel model.]{\includegraphics[width=0.40\linewidth, height=0.135\linewidth]{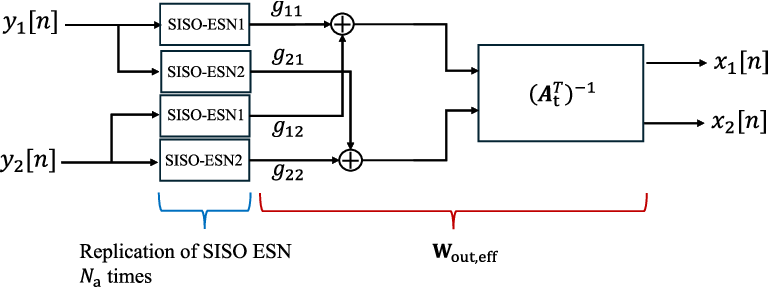}\label{fig:parametric_mimo_2x2_fig3}}%
    \caption{ESN configuration for a parametric $2 \times 2$ MIMO channel with $N_{\mathrm{path}} = 2$ paths with \emph{distinct} PDP per delay tap: Steps 2 and 3.}
\end{figure*}

Consider the scenario $N_{\mathrm{t}} = N_{\mathrm{r}}$. 
The transfer function of the equalizer of the MIMO channel can be expressed as $\mathbf{H}_{\mathrm{eq}}(z) = 
(\boldsymbol{A}_{\mathrm{t}}^{T})^{\dagger}
\left( \mathbf{D}(z) \right)^{-1}
\boldsymbol{A}_{\mathrm{r}}^{\dagger}  $.
Similar to the analysis of Sec.~\ref{sec:factorizable_mimo_channel}, let $N_{\mathrm{t}} = 2$ and also let $N_{\mathrm{p}} = 2$.  
First, we consider the most general case where the $N_{\mathrm{p}} = 2$ paths represent two \emph{distinct} power delay profiles (PDPs) and thus, two distinct SISO channel statistics induced by $h_1(z) \overset{\mtrian}{=} \sum_{\ell=0}^{L-1} c_0^{(\ell)} z^{-\ell}$ and $h_2(z) \overset{\mtrian}{=} \sum_{\ell=0}^{L-1} c_1^{(\ell)} z^{-\ell}$. 
The equalizer structure represented by $\mathbf{H}_{\mathrm{eq}}(z)$ for $N_{\mathrm{t}} = N_{\mathrm{p}} = 2$ is depicted in Fig.~\ref{fig:parametric_mimo_2x2_fig1}. 
Since the motivation is to move all the trainable weights to the output, the weights matrix corresponding to $\boldsymbol{A}_{\mathrm{r}}^{-1}$ can be moved to the right of the parallel branches represented by the SISO ESN equalizers $\sfrac{1}{h_1(z)}$ and $\sfrac{1}{h_2(z)}$. 
However, unlike the case in Sec.~\ref{sec:factorizable_mimo_channel}, there is a repetition due to the movement of $\boldsymbol{A}_{\mathrm{r}}^{-1}$to the right, when $h_1(z)$ and $h_2(z)$ have distinct PDPs with distinct channel statistics.
This results in the transfer functions of each of the $N_{\mathrm{p}}$ distinct SISO ESN equalizers being repeated $N_{\mathrm{t}}$ times, while $\boldsymbol{A}_{\mathrm{r}}^{-1},$ is absorbed into the effective output weights along with $\boldsymbol{A}_{\mathrm{t}}^{-1}$.
This sequence of steps is depicted visually in Fig.~\ref{fig:parametric_mimo_2x2_fig2} and Fig.~\ref{fig:parametric_mimo_2x2_fig3}.
Similar to Sec.~\ref{sec:factorizable_mimo_channel}, the MIMO ESN equalizer can be decomposed as two separate SISO ESNs. 
The configured weights $\mathbf{W}_{\mathrm{res}}$ and $\mathbf{W}_{\mathrm{in}}$ can be written as $\mathbf{W}_{\mathrm{res,eff}} = \blkdiag\big(\mathbf{W}_{\mathrm{res,1}}, \mathbf{W}_{\mathrm{res,2}}, \mathbf{W}_{\mathrm{res,1}}, \mathbf{W}_{\mathrm{res,2}}\big)$ and $\mathbf{W}_{\mathrm{in,eff}} = \blkdiag\big(\mathbf{w}_{\mathrm{in,1}}, \mathbf{w}_{\mathrm{in,2}}, \mathbf{w}_{\mathrm{in,1}}, \mathbf{w}_{\mathrm{in,2}} \big)$, where $\{ \mathbf{W}_{\mathrm{res},i}, \mathbf{w}_{\mathrm{in},i} \}$ for the $i^{\text{th}}$ SISO ESN are configured using the statistics of $h_{i}(z)$ for $i=1,2$.
The addition of input windowing as well as output skip and delayed connections with window length $N_w$ will result in the WESN weight configuration architecture for MIMO systems.
If the number of reservoir/RNN neurons in each SISO ESN is $N_{\mathrm{n}}$ and each of the $N_{\mathrm{p}}$ paths obey distinct statistics, the number of reservoir/RNN neurons in the configured WESN scales as $N_{\mathrm{n, MIMO}} = N_{\mathrm{t}} N_{\mathrm{p}} N_{\mathrm{n}}$ for MIMO systems.
This architecture and scaling are primarily due to the statistically distinct PDPs.

In 3GPP/ITU MIMO channels such as the SCM~\cite{3gppTR25996}, the combination of the $N_{\mathrm{p}}$ paths results in an average PDP that is tailored to a specific environment, e.g., suburban macro, urban micro, etc. 
For example, Section 5.4 in~\cite{3gppTR25996} specifies MIMO channels through the PDP and the AoA/AoD-related information for various environments.
The environment-specific average PDP together with the AoAs/AoD-related information can be estimated~\cite{RubayetDoA} allowing us to empirically obtain the statistics for the underlying SISO channel to conduct reservoir/RNN weight configuration following the procedure outlined in Sec.~\ref{sec:factorizable_mimo_channel}.
Accordingly, the architecture of the configured WESN for general MIMO systems can be simplified, resulting in $N_{\mathrm{n, MIMO}} = N_{\mathrm{t}} N_{\mathrm{n}}$.

\section{Performance Evaluation}
\label{sec:performance_evaluation}

\subsection{Experimental Settings}
\begin{figure*}[ht]
    \centering    
    \subfloat[Conventional approaches]{\includegraphics[width=0.6075\linewidth]{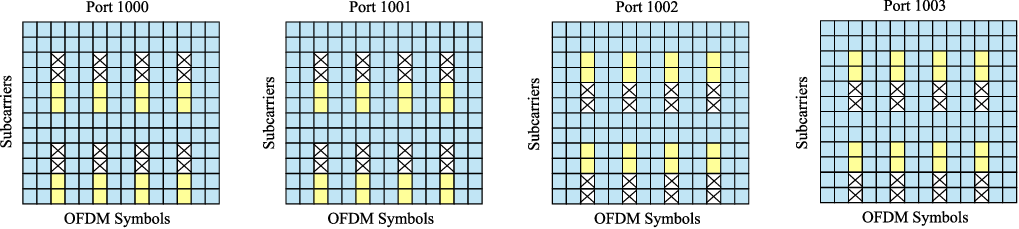}\label{fig:mimo_scattered_pilot_pattern_orthogonal}} \\
    \subfloat[Learning-based (RC-based) approaches]{\includegraphics[width=0.6075\linewidth]{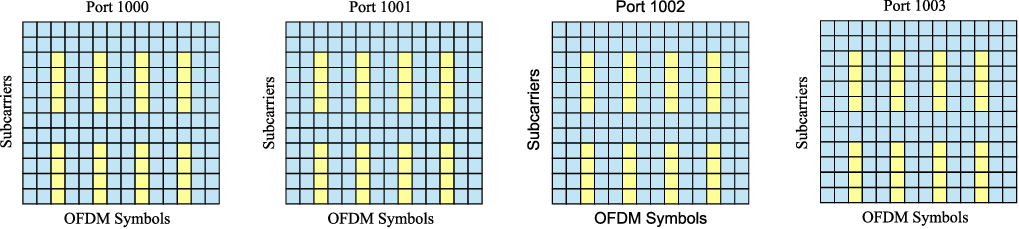}\label{fig:mimo_scattered_pilot_pattern_non-orthogonal}}%
    \caption{MIMO scattered (comb) RS pattern ($N_{\mathrm{t}} = N_{\mathrm{r}} = 4$) in one resource block (RB) for conventional and RC-based symbol detection approaches.}
    \label{fig:mimo_scattered_pilot_pattern}
\end{figure*}

The effectiveness of reservoir/RNN weight configuration procedures for WESN are verified through extensive simulations for both OFDM and $4 \times 4$ MIMO-OFDM systems.
The number of subcarriers of the underlying OFDM system is chosen to be $N_{\mathrm{sc}} = 1024$ and the cyclic prefix (CP) length to be $N_{\mathrm{cp}} = 160$.
We consider the clustered delay line (CDL) channel model~\cite{std3gpp38901}, specifically CDL-D and CDL-E, as well as the SCM~\cite{3gppTR25996} in our evaluations. 
The user speed is set to $5$ km/hr with carrier frequency $f_c = 3.5$ GHz, corresponding to band n$78$ in 5G NR.
In addition, the scattered (comb) RS pattern is adopted, as specified in 3GPP 5G New Radio (NR) standards~\cite{std3gpp38211, std3gpp38212}.
The comb RS pattern adopted for the $4 \times 4$ MIMO-OFDM system is shown in Fig.~\ref{fig:mimo_scattered_pilot_pattern}, where the RS resource elements (REs) are indicated in yellow and the data REs are indicated in blue.
The white REs with cross markers indicate empty RS REs. 
The baseline symbol detection approach considered for comparison is linear minimum mean square error (LMMSE)-based symbol detection using estimated channel state information (CSI) that is first computed at the RS RE locations and then interpolated to all RE locations via two-dimensional LMMSE interpolation (2D-LMMSE interpolation)~\cite{Dong2007}.
Furthermore, we consider line-of-sight (LOS) channels in the evaluations, encompassing individual realizations that may be strictly MP or mixed-phase.
Consideration of non-line-of-sight (NLOS) channels, including those with a dominant strictly NMP factor, potentially requires configuring or fine-tuning the feedforward output weights of the WESN in addition to online OTA RS-based training, which will be considered in future work. 
The WESN implementations employ the hyperbolic tangent (tanh) activation.
For the randomly generated WESN, the spectral radius is set to $0.4$ and the reservoir sparsity to $0.6$.
Delay learning~\cite{zhou2019}, which is theoretically justified in Sec.~\ref{sec:connection_with_wesn}, is also implemented.

\subsection{OFDM Symbol Detection: Strictly MP Channel}
\label{sec:performance_evaluation_strictly_MP_SISO}
\begin{figure}[!h]
    \centering    \includegraphics[width=0.752\linewidth]{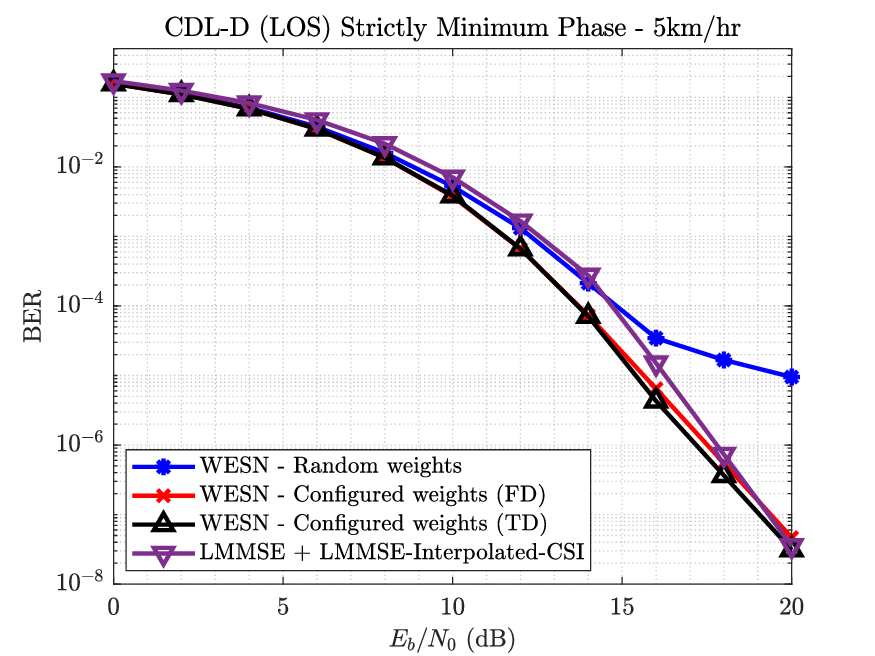}
    \caption{BER performance with 16-QAM for WESN in the strictly MP CDL-D channel. `FD': `frequency-domain', `TD': `time-domain'.}
    \label{fig:BER_SISO_CDL-D_MP}
\end{figure}

In this section, we validate the effectiveness of the frequency/time-domain reservoir/RNN weight configuration procedures for WESN under strictly MP channels, where the CDL-D channel~\cite{std3gpp38901} is used for this evaluation. 
The WESN configured using the frequency-domain method uses a total of $ML_{\mathrm{rp}}$ reservoir neurons with $M=5$ significant eigenvectors and denominator order $L_{\mathrm{rp}}=7$ in the rational polynomial approximation, resulting in a total of $35$ reservoir/RNN neurons.
Similarly, for the WESN configured using the time-domain method, we use $M=5$ significant eigenvectors and cut-off point $L_{\mathrm{f}}=7$ for the reduced-order IIR filter, also resulting in $M L_{\mathrm{f}} = 35$ reservoir/RNN recurrent neurons.
For the WESN with randomly generated weights as well as the configured WESNs, a window length of $N_w = 5$ is used, resulting in $L_{ff} = 5$ feedforward taps in the WESN, following the description in Sec.~\ref{sec:connection_with_wesn}.
From Fig.~\ref{fig:BER_SISO_CDL-D_MP}, we can see that for the strictly MP CDL-D channel, both the frequency/time-domain  configuration methods provide near identical level of BER performance improvement compared to randomly generated input and reservoir/RNN weights, thus validating the time-domain configuration procedure.
Also, the WESN with randomly generated weights exhibits an error floor behavior in the high signal-to-noise ratio (SNR) regime, while weight configuration can effectively mitigate this issue.

\subsection{OFDM Symbol Detection: Mixed-Phase Channel}
\begin{figure}[htbp]
    \centering    \includegraphics[width=0.752\linewidth]{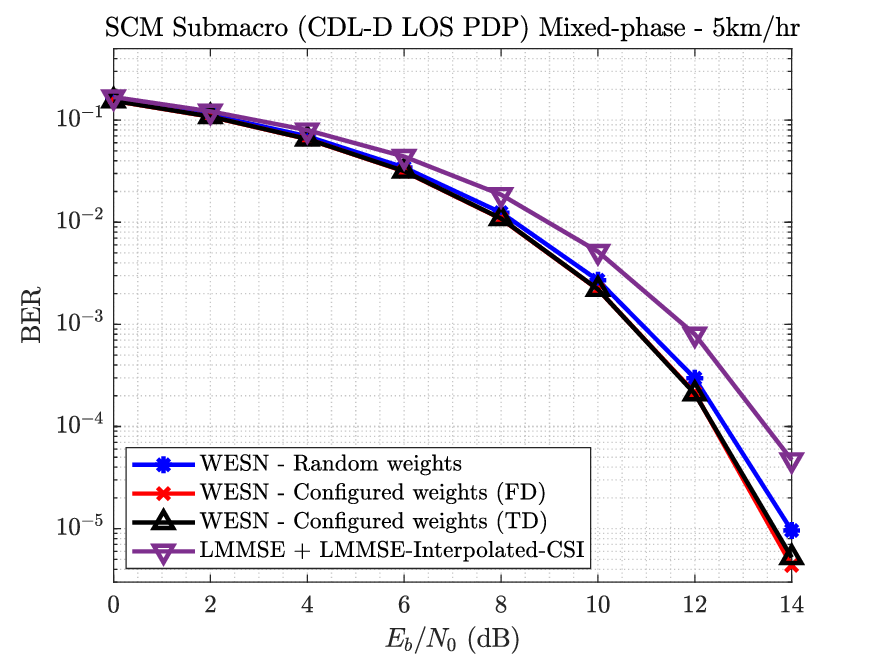}
    \caption{BER performance with 16-QAM for WESN for a mixed-phase SCM channel with a CDL-D LOS power delay profile (PDP). `FD': `frequency-domain', `TD': `time-domain'.}
    \label{fig:BER_SISO_CDL-D_mixed_phase}
\end{figure}

This section validates the effectiveness of reservoir/RNN weight configuration procedures for WESN under mixed-phase channels.
Specifically, we utilize the CDL-D PDP in the 3GPP SCM~\cite{3gppTR25996}, resulting in individual channel realizations that are mixed-phase.
A total of $35$ reservoir/RNN neurons are employed corresponding to $M=5$ and $L_{\mathrm{f}} = 7$ (time-domain method) and $L_{\mathrm{rp}}=7$ (frequency-domain method) and a window length of $N_w = 5$.
The BER performance with configured weights using both procedures is shown in Fig.~\ref{fig:BER_SISO_CDL-D_mixed_phase}. 
We can see that for mixed-phase channels both frequency/time-domain configuration methods utilizing the strictly MP factor lead to similar performance improvement compared to randomly generated weights.
Furthermore, the performance improvement with configured weights is smaller compared to that for strictly MP channels shown in Sec.~\ref{sec:performance_evaluation_strictly_MP_SISO}.
This points towards a potential requirement of configuration or fine-tuning of the feedforward weights in the WESN to further assist equalization of the underlying strictly NMP component of the mixed-phase channel.

\subsection{MIMO-OFDM Symbol Detection}
\label{sec:performance_evaluation_mimo_parametric_scm}
\begin{figure}[htbp]
    \centering    \includegraphics[width=0.752\linewidth]{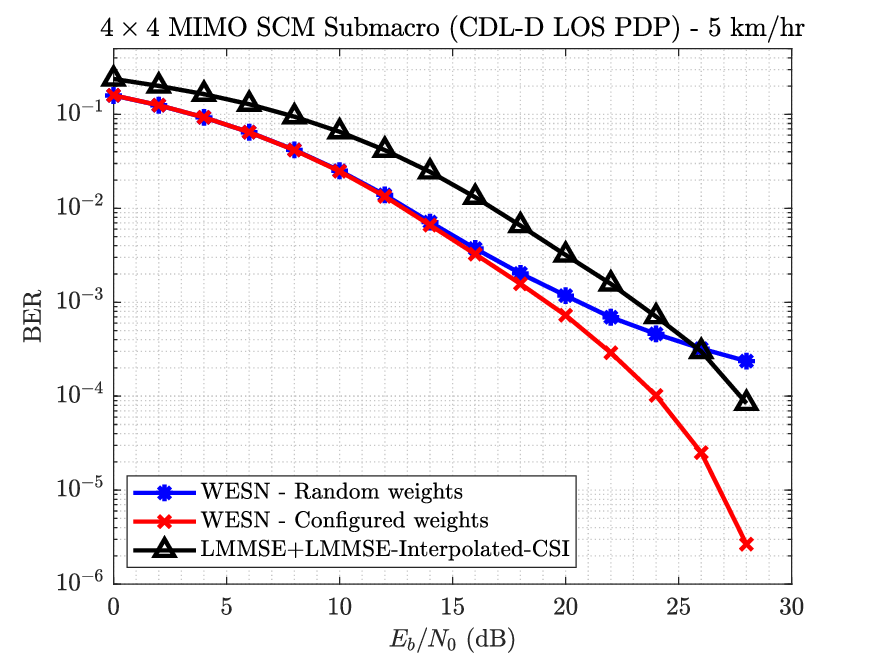}
    \caption{BER performance with 16-QAM for WESN for the $4 \times 4$ MIMO SCM channel model with a CDL-D LOS power delay profile (PDP).}
    \label{fig:BER_MIMO_4x4_SCM_Submacro_CDL-D_PDP}
\end{figure}

\begin{figure}[htbp]
    \centering    \includegraphics[width=0.752\linewidth]{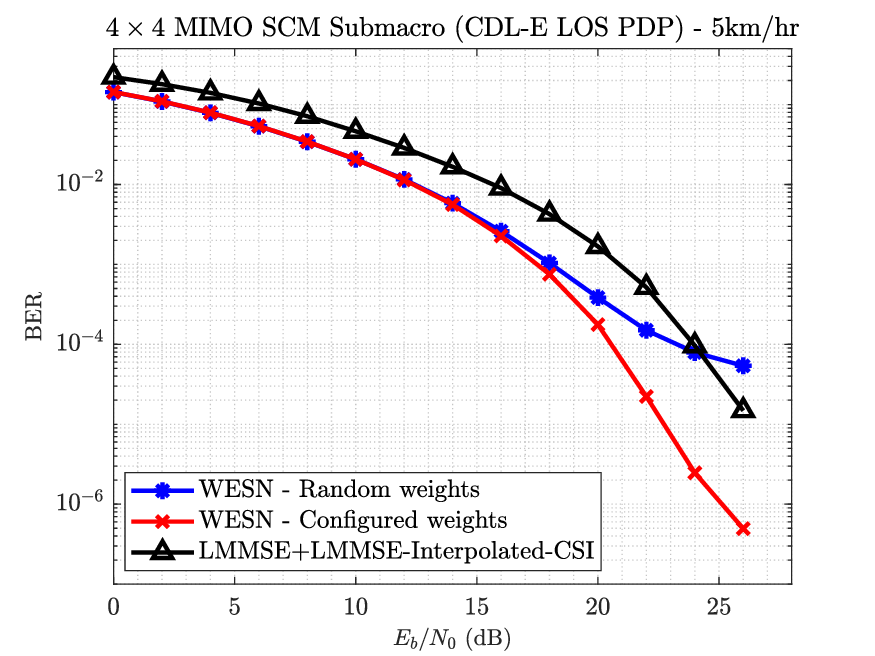}
    \caption{BER performance with 16-QAM for WESN for the $4 \times 4$ MIMO SCM channel model with a CDL-E LOS power delay profile (PDP).}
    \label{fig:BER_MIMO_4x4_SCM_Submacro_CDL-E_PDP}
\end{figure}

We consider a $4 \times 4$ MIMO system with ULA antenna geometry at both the transmitter and 
the receiver under the 3GPP SCM~\cite{3gppTR25996}.
We use the suburban macro scenario within the SCM channel model, where the distribution parameters for the AoAs and AoDs of the individual paths at each delay tap are set following Tables 5.1, 5.2 and 5.3 in~\cite{3gppTR25996}.
In order to emphasize LOS scenarios, we consider the CDL-D and CDL-E LOS PDPs in this evaluation.
The symbol detection performance for $4\times 4$ MIMO under the CDL-D and CDL-E PDPs are shown in Fig.~\ref{fig:BER_MIMO_4x4_SCM_Submacro_CDL-D_PDP} and Fig.~\ref{fig:BER_MIMO_4x4_SCM_Submacro_CDL-E_PDP} respectively, where the frequency-domain method is used to configure $\mathbf{W}_{\mathrm{res}}$ and $\mathbf{w}_{\mathrm{in}}$ of each SISO ESN. 
For each SISO ESN, we use $M=3$ eigenvectors and $L_{\mathrm{rp}}=3$-rd order rational polynomial resulting in $M L_{\mathrm{rp}} = 9$ neurons. 
The SISO ESN is replicated $N_{\mathrm{t}} = 4$ times, resulting in a total of $36$ reservoir/RNN recurrent neurons in the MIMO ESN.
A window length of $N_w = 5$ is used to reach the final WESN architecture.
Fig.~\ref{fig:BER_MIMO_4x4_SCM_Submacro_CDL-D_PDP} and Fig.~\ref{fig:BER_MIMO_4x4_SCM_Submacro_CDL-E_PDP} demonstrate that configured weights greatly aid in mitigating the error floor issue at high SNR that is characteristic of the WESN with randomly generated input and reservoir/RNN weights, especially in the MIMO setting. 

\section{Conclusion}
\label{sec:conclusion}

In this work, we present a principled understanding of the ESN, a popular architecture in the RC paradigm. 
Building on prior work, a fundamental theoretical analysis of the vanilla ESN as well as the WESN architecture is developed grounded in signal processing ideas.
It provides valuable insights into the reasons behind their effectiveness in the symbol detection task for both OFDM and MIMO-OFDM systems. 
Furthermore, systematic approaches are developed to establish a foundational linkage between available domain knowledge in the form of channel statistics and direct configuration of the untrained reservoir/RNN weights of the WESN.
The introduced weight configuration procedures are validated under both OFDM and MIMO-OFDM symbol detection tasks through extensive simulations, demonstrating significant performance improvement.
This paves the path for explainable AI/xAI in wireless applications, especially in real-time PHY processing.
We believe this is an important step towards systematically incorporating domain knowledge into the design of neural networks for engineering applications.

\section*{Acknowledgment}  
This material is based upon work supported by the National Science Foundation under grants 2002908, 2003059, 2148212 and is supported in part by funds from federal agency and industry partners as specified in the Resilient \& Intelligent NextG Systems (RINGS) program.
The authors would also like to thank Dr. Robert Calderbank and Ibraheem Alturki for insightful discussions.

\begin{appendices}

\section{Proof of Lemma~\ref{lemma:pca_min_error}}
\label{sec:Appendix_A}
Let $\mathbf{g} \in \mathbb{C}^{N}$ denote the impulse response of the time-domain equalizer for the channel impulse response represented by $[\mathbf{h}^T, \, \boldsymbol{0}_{(N-L) \times 1}^T]^T \in \mathbb{C}^N$.
Analyzing the objective of Eq.~\eqref{eq:pca_optimization_time_domain}:
\begin{align}
    \mathbb{E}_{\mathbf{g}} \left[\|\mathbf{F}\mathbf{F}^H \mathbf{g} - \mathbf{g}\|_2^2\right],
    &=\mathbb{E}_{\mathbf{g}} \left[\Tr\left((\mathbf{F}\mathbf{F}^H \mathbf{g} - \mathbf{g})(\mathbf{F}\mathbf{F}^H \mathbf{g} - \mathbf{g})^H\right)\right], \nonumber \\
    &= \mathbb{E}_{\mathbf{g}} \left[\Tr\left((\mathbf{F}\mathbf{F}^H - \mathbf{I})\mathbf{g}\mathbf{g}^H(\mathbf{F}\mathbf{F}^H - \mathbf{I})\right)\right] \nonumber \\
    &= \Tr\left((\mathbf{F}\mathbf{F}^H - \mathbf{I})\mathbb{E}_{\mathbf{g}}[\mathbf{g}\mathbf{g}^H](\mathbf{F}\mathbf{F}^H - \mathbf{I})\right).
    \label{eq:appendix_A_step1}
\end{align}
Denote by $\mathbf{K} \in \mathbb{C}^{N \times N}$ the covariance matrix of $\mathbf{g}$, i.e.,  $\mathbf{K} \overset{\mtrian}{=} \mathbb{E}_{\mathbf{g}}[\mathbf{g}\mathbf{g}^H]$.
Its eigen-decomposition is obtained as 
$\mathbf{K} = \mathbf{V} \mathbf{\Lambda} \mathbf{V}^H$, where the eigenvectors are contained in the columns of $\mathbf{V} \in \mathbb{C}^{N \times N}$ and $\mathbf{\Lambda} = \diag\left(\{\lambda_i \}_{i=0}^{N-1} \right) \in \mathbb{R}^{N \times N}$ contains the corresponding eigenvalues. 
Substituting for \(\mathbf{K}\) in Eq.~\eqref{eq:appendix_A_step1}, 
\begin{align}
    &\mathbb{E}_{\mathbf{g}} \left[\|\mathbf{F}\mathbf{F}^H \mathbf{g} - \mathbf{g}\|_2^2\right] \nonumber \\
    &=\Tr(\mathbf{F}\mathbf{F}^H\mathbf{K}\mathbf{F}\mathbf{F}^H - \mathbf{F}\mathbf{F}^H\mathbf{K} - \mathbf{K}\mathbf{F}\mathbf{F}^H + \mathbf{K}).
    \label{eq:appendix_A_step2}
\end{align}
Next, the first three terms inside the trace in Eq.~\eqref{eq:appendix_A_step2} can be simplified as \(\mathbf{V} \mathbf{\Lambda}_M \mathbf{V}^H\), where $\mathbf{\Lambda}_M = \diag\left(\{\lambda_i \}_{i=0}^{M-1} \right) \in \mathbb{R}^{M \times M}$ is the truncated matrix containing only the first \(M\) most significant eigenvalues. 
Combining the simplified terms, we get $\mathbb{E}_{\mathbf{g}}\left[\|\mathbf{F}\mathbf{F}^H \mathbf{g} - \mathbf{g}\|_2^2\right] = \Tr(\mathbf{V} \mathbf{\Lambda} \mathbf{V}^H) - \Tr \left(\mathbf{V} \blkdiag(\mathbf{\Lambda}_M, \boldsymbol{0}_{(N-M) \times (N-M)}) \mathbf{V}^H \right) = \sum_{j=M}^{N-1} \lambda_j$, thus concluding the proof of Lemma~\ref{lemma:pca_min_error}.

\section{Proof of Theorem~\ref{theorem:bound_on_error}}
\label{sec:Appendix_B}

Define the lower shift matrix $\boldsymbol{L}_i \in \mathbb{R}^{N \times N}$ as $\boldsymbol{L}_i \overset{\mtrian}{=} [\boldsymbol{0}, \boldsymbol{0}; \mathbf{I}_{N-i}, \boldsymbol{0}]$, 
resulting in
$\boldsymbol{L}_i^H \boldsymbol{L}_i = [\mathbf{I}_{N-i}, \boldsymbol{0}; \boldsymbol{0}, \boldsymbol{0}]$.
We use the notation $\mathbf{F} \overset{\mtrian}{=} \mathbf{F}_{\mathrm{opt}}^{(\mathrm{P}^{*})} = [\mathbf{V}]_{:,0:M-1}$.
Defining $\widetilde{\mathbf{G}} \overset{\mtrian}{=} \widetilde{\mathbf{H}}^{-1}$,
the objective of $\textbf{P2}$ \eqref{eq:F_P1} evaluated at $\mathbf{F}_{\mathrm{opt}}^{(\mathrm{P}^{*})}$ is
\begin{align}
    \varepsilon_{\mathrm{app}}^{(P2)} = \mathbb{E} \left[ \| \mathbf{F} \mathbf{F}^{H} \widetilde{\mathbf{G}}  - \widetilde{\mathbf{G}}    \|_F^2 \right] = \sum_{i=0}^{N-1} \mathbb{E}_{\mathbf{g}} \left[ \| \mathbf{F} \mathbf{F}^H \boldsymbol{L}_i \mathbf{g} - \boldsymbol{L}_i \mathbf{g} \|_2^2 \right].
    \label{eq:P2_evaluated_at_F_PCA}
\end{align}
Next, we closely examine the term $ \| \mathbf{F} \mathbf{F}^H \boldsymbol{L}_i \mathbf{g} - \boldsymbol{L}_i \mathbf{g} \|_2^2$:
\begin{align}
    &\| \mathbf{F} \mathbf{F}^H \boldsymbol{L}_i \mathbf{g} - \boldsymbol{L}_i \mathbf{g} \|_2^2 = (\mathbf{F} \mathbf{F}^H \boldsymbol{L}_i \mathbf{g} - \boldsymbol{L}_i \mathbf{g})^H (\mathbf{F} \mathbf{F}^H \boldsymbol{L}_i \mathbf{g} - \boldsymbol{L}_i \mathbf{g}), \nonumber \\
    &= \mathbf{g}^H \boldsymbol{L}_i^H \boldsymbol{L}_i \mathbf{g} - \mathbf{g}^H \boldsymbol{L}_i^H \mathbf{F} \mathbf{F}^H \boldsymbol{L}_i \mathbf{g}, \nonumber \\
    & \overset{(a)}{=}\Tr(\mathbf{g} \mathbf{g}^H \boldsymbol{L}_i^H \boldsymbol{L}_i) - \Tr(\mathbf{g} \mathbf{g}^H \boldsymbol{L}_i^H \mathbf{F} \mathbf{F}^H \boldsymbol{L}_i).
\end{align}
where $(a)$ holds due to the application of the trace and its cyclic shift property.
It follows that
\begin{align}
    &\mathbb{E}_{\mathbf{g}} \left[ \| \mathbf{F} \mathbf{F}^H \boldsymbol{L}_i \mathbf{g} - \boldsymbol{L}_i \mathbf{g} \|_2^2 \right]  \nonumber \\ &=\Tr\left(\mathbb{E}_{\mathbf{g}}[\mathbf{g} \mathbf{g}^H] \boldsymbol{L}_i^H \boldsymbol{L}_i \right) - \Tr \left(\mathbb{E}_{\mathbf{g}}[\mathbf{g} \mathbf{g}^H ]\boldsymbol{L}_i^H \mathbf{F} \mathbf{F}^H \boldsymbol{L}_i \right), \nonumber \\
    &= \Tr(\mathbf{K} \boldsymbol{L}_i^H \boldsymbol{L}_i ) - \Tr(\mathbf{K} \boldsymbol{L}_i^H \mathbf{F}\mathbf{F}^H \boldsymbol{L}_i).
    \label{eq:difference_of_traces}
\end{align}
Therefore, the final approximation error is given by
\begin{align}
    \varepsilon_{\mathrm{app}}^{(\mathrm{P2})} = \sum_{i=0}^{N-1} [\Tr(\mathbf{K} \boldsymbol{L}_i^H \boldsymbol{L}_i ) - \Tr(\mathbf{K} \boldsymbol{L}_i^H \mathbf{F}\mathbf{F}^H \boldsymbol{L}_i)].
\end{align}
Note that $\mathbf{K} \boldsymbol{L}_i^H \boldsymbol{L}_i$ and $\mathbf{K} \boldsymbol{L}_i^H \mathbf{F}\mathbf{F}^H \boldsymbol{L}_i$ are not strict principal submatrices of $\mathbf{K}$. 
Therefore, exact expressions or upper bounds on their traces in terms of the eigenvalues of $\mathbf{K}$ are difficult to derive using the eigenvalue interlacing theorem. 
This concludes the proof of Theorem~\ref{theorem:bound_on_error}.

\end{appendices}

\bibliographystyle{IEEEtran}
\bibliography{IEEEabrv,ref}

% Generated by IEEEtran.bst, version: 1.14 (2015/08/26)
\begin{thebibliography}{10}
\providecommand{\url}[1]{#1}
\csname url@samestyle\endcsname
\providecommand{\newblock}{\relax}
\providecommand{\bibinfo}[2]{#2}
\providecommand{\BIBentrySTDinterwordspacing}{\spaceskip=0pt\relax}
\providecommand{\BIBentryALTinterwordstretchfactor}{4}
\providecommand{\BIBentryALTinterwordspacing}{\spaceskip=\fontdimen2\font plus
\BIBentryALTinterwordstretchfactor\fontdimen3\font minus \fontdimen4\font\relax}
\providecommand{\BIBforeignlanguage}[2]{{%
\expandafter\ifx\csname l@#1\endcsname\relax
\typeout{** WARNING: IEEEtran.bst: No hyphenation pattern has been}%
\typeout{** loaded for the language `#1'. Using the pattern for}%
\typeout{** the default language instead.}%
\else
\language=\csname l@#1\endcsname
\fi
#2}}
\providecommand{\BIBdecl}{\relax}
\BIBdecl

\bibitem{JereMILCOM2023}
S.~Jere, K.~Said, L.~Zheng, and L.~Liu, ``{Towards Explainable Machine Learning: The Effectiveness of Reservoir Computing in Wireless Receive Processing},'' in \emph{2023 IEEE Military Commun. Conf. (MILCOM)}, 2023, pp. 667--672.

\bibitem{Shafin2020}
R.~Shafin, L.~Liu, V.~Chandrasekhar, H.~Chen, J.~Reed, and J.~Zhang, ``Artificial {I}ntelligence-{E}nabled {C}ellular {N}etworks: {A} {C}ritical {P}ath to beyond-{5G} and {6G},'' \emph{IEEE Wireless Commun. Mag.}, vol.~27, no.~2, pp. 212--217, 2020.

\bibitem{Xu2024}
J.~Xu, S.~Jere, Y.~Song, Y.-H. Kao, L.~Zheng, and L.~Liu, ``{Learning at the Speed of Wireless: Online Real-Time Learning for AI-Enabled MIMO in NextG},'' \emph{{IEEE} Commun. Mag.}, pp. 1--7, 2024.

\bibitem{Lukosevicius2012}
M.~Luko{\v{s}}evi{\v{c}}ius, \emph{A Practical Guide to Applying Echo State Networks}.\hskip 1em plus 0.5em minus 0.4em\relax Springer Berlin Heidelberg, 2012, pp. 659--686.

\bibitem{zhou2019}
Z.~Zhou, L.~Liu, and H.-H. Chang, ``Learning for {D}etection: {MIMO-OFDM} {S}ymbol {D}etection {T}hrough {D}ownlink {P}ilots,'' \emph{{IEEE} Trans. Wireless Commun.}, vol.~19, no.~6, pp. 3712--3726, 2020.

\bibitem{zhou2020rcnet}
Z.~Zhou, L.~Liu, S.~Jere, J.~Zhang, and Y.~Yi, ``{RCNet}: {I}ncorporating {S}tructural {I}nformation {I}nto {D}eep {RNN} for {O}nline {MIMO-OFDM} {S}ymbol {D}etection {W}ith {L}imited {T}raining,'' \emph{{IEEE} Trans. Wireless Commun.}, vol.~20, no.~6, pp. 3524--3537, 2021.

\bibitem{RCstruct}
J.~Xu, Z.~Zhou, L.~Li, L.~Zheng, and L.~Liu, ``{RC-Struct: A Structure-Based Neural Network Approach for MIMO-OFDM Detection},'' \emph{{IEEE} Trans. Wireless Commun.}, vol.~21, no.~9, pp. 7181--7193, 2022.

\bibitem{LiuAI2}
H.-H. Chang, H.~Song, Y.~Yi, J.~Zhang, H.~He, and L.~Liu, ``{Distributive Dynamic Spectrum Access Through Deep Reinforcement Learning: A Reservoir Computing-Based Approach},'' \emph{IEEE Internet Things J.}, vol.~6, no.~2, pp. 1938--1948, 2019.

\bibitem{Chang2020}
H.-H. Chang, L.~Liu, and Y.~Yi, ``Deep {E}cho {S}tate {Q}-network ({DEQN}) and {I}ts {A}pplication in {D}ynamic {S}pectrum {S}haring for 5{G} and {B}eyond,'' \emph{IEEE Trans. Neur. Netw. Learn. Syst.}, vol.~33, no.~3, pp. 929--939, 2022.

\bibitem{Khani2020}
M.~Khani, M.~Alizadeh, J.~Hoydis, and P.~Fleming, ``Adaptive {N}eural {S}ignal {D}etection for {M}assive {MIMO},'' \emph{{IEEE} Trans. Wireless Commun.}, vol.~19, no.~8, pp. 5635--5648, 2020.

\bibitem{DSAComparison}
H.~Mosavat-Jahromi, Y.~Li, L.~Cai, and J.~Pan, ``{Prediction and Modeling of Spectrum Occupancy for Dynamic Spectrum Access Systems},'' \emph{IEEE Trans. Cogn. Commun. Netw.}, vol.~7, no.~3, pp. 715--728, 2021.

\bibitem{samuel2019learning}
N.~Samuel, T.~Diskin, and A.~Wiesel, ``Learning to {D}etect,'' \emph{{IEEE} Trans. Signal Process.}, vol.~67, no.~10, pp. 2554--2564, 2019.

\bibitem{OAMPNet2018}
H.~He, C.-K. Wen, S.~Jin, and G.~Y. Li, ``A {M}odel-{D}riven {D}eep {L}earning {N}etwork for {MIMO} {D}etection,'' in \emph{2018 IEEE Global Conf. on Sig. and Inf. Proc. (GlobalSIP)}, 2018, pp. 584--588.

\bibitem{Goutay2020}
M.~Goutay, F.~Ait~Aoudia, and J.~Hoydis, ``Deep {H}yper{N}etwork-{B}ased {MIMO} {D}etection,'' in \emph{2020 IEEE 21st Intl. Workshop on Sig. Proc. Adv. in Wireless Commun. (SPAWC)}, 2020, pp. 1--5.

\bibitem{mosleh2017brain}
S.~S. Mosleh, L.~Liu, C.~Sahin, Y.~R. Zheng, and Y.~Yi, ``Brain-{I}nspired {W}ireless {C}ommunications: {W}here {R}eservoir {C}omputing {M}eets {MIMO-OFDM},'' \emph{{IEEE} Trans. Neural Netw. Learn. Syst.}, vol.~29, no.~10, pp. 4694--4708, 2018.

\bibitem{Montavon2019}
G.~Montavon, A.~Binder, S.~Lapuschkin, W.~Samek, and K.-R. M{\"u}ller, \emph{Layer-Wise Relevance Propagation: An Overview}.\hskip 1em plus 0.5em minus 0.4em\relax Cham: Springer International Publishing, 2019, pp. 193--209.

\bibitem{Ozturk2007}
M.~C. Ozturk, D.~Xu, and J.~C. Pr\'{\i}ncipe, ``{Analysis and Design of Echo State Networks},'' \emph{Neural Comput.}, vol.~19, no.~1, p. 111–138, 2007.

\bibitem{Bollt2021}
E.~Bollt, ``On explaining the surprising success of reservoir computing forecaster of chaos? {T}he universal machine learning dynamical system with contrast to {VAR} and {DMD},'' \emph{Chaos}, vol.~31, p. 013108, 2021.

\bibitem{Halus2019}
A.~Haluszczynski and C.~Räth, ``Good and bad predictions: Assessing and improving the replication of chaotic attractors by means of reservoir computing,'' \emph{Chaos}, vol.~29, no.~10, p. 103143, 2019.

\bibitem{Carroll2022}
T.~L. Carroll, ``Optimizing memory in reservoir computers,'' \emph{Chaos}, vol.~32, no.~2, p. 023123, 2022.

\bibitem{JereTCOM2023}
S.~Jere, R.~Safavinejad, and L.~Liu, ``{Theoretical Foundation and Design Guideline for Reservoir Computing-based MIMO-OFDM Symbol Detection},'' \emph{{IEEE} Trans. Commun.}, vol.~71, no.~9, pp. 5169--5181, 2023.

\bibitem{Gonon2020}
L.~Gonon, L.~Grigoryeva, and J.-P. Ortega, ``{R}isk {B}ounds for {R}eservoir {C}omputing,'' \emph{Jour. Mach. Learn. Res.}, vol.~21, no. 240, pp. 1--61, 2020.

\bibitem{Jere2023WCL}
S.~Jere, R.~Safavinejad, L.~Zheng, and L.~Liu, ``{Channel Equalization Through Reservoir Computing: A Theoretical Perspective},'' \emph{IEEE Wireless Communications Letters}, vol.~12, no.~5, pp. 774--778, 2023.

\bibitem{JereJSTSP2024}
S.~Jere, L.~Zheng, K.~Said, and L.~Liu, ``{Universal Approximation of Linear Time-Invariant (LTI) Systems Through RNNs: Power of Randomness in Reservoir Computing},'' \emph{{IEEE} J. Sel. Topics Signal Process.}, vol.~18, no.~2, pp. 184--198, 2024.

\bibitem{oppenheim2009discrete}
A.~V. Oppenheim, R.~W. Schafer, and J.~R. Buck, \emph{Discrete-Time Signal Processing}, 3rd~ed.\hskip 1em plus 0.5em minus 0.4em\relax Upper Saddle River, NJ: Prentice Hall, 2009.

\bibitem{Vaidyanathan1993}
P.~P. Vaidyanathan, \emph{Multirate Systems and Filter Banks}.\hskip 1em plus 0.5em minus 0.4em\relax USA: Prentice-Hall, Inc., 1993.

\bibitem{Zhu2017}
Z.~Zhu and M.~B. Wakin, ``{On the Asymptotic Equivalence of Circulant and Toeplitz Matrices},'' \emph{{IEEE} Trans. Inf. Theory}, vol.~63, no.~5, pp. 2975--2992, 2017.

\bibitem{Moir2022}
\BIBentryALTinterwordspacing
T.~J. Moir, \emph{Toeplitz Convolution Matrix Method}.\hskip 1em plus 0.5em minus 0.4em\relax Cham: Springer International Publishing, 2022, pp. 279--295. [Online]. Available: \url{https://doi.org/10.1007/978-3-030-76947-5_10}
\BIBentrySTDinterwordspacing

\bibitem{3gppTR25996}
\emph{Universal Mobile Telecommunications System (UMTS); Spatial channel model for Multiple Input Multiple Output (MIMO) simulations}, 3GPP Std. TR 25.996, Rev. 16.0.0, 2020.

\bibitem{RubayetDoA}
R.~Shafin, L.~Liu, Y.~Li, A.~Wang, and J.~Zhang, ``{Angle and Delay Estimation for 3-D Massive MIMO/FD-MIMO Systems Based on Parametric Channel Modeling},'' \emph{{IEEE} Trans. Wireless Commun.}, vol.~16, no.~8, pp. 5370--5383, 2017.

\bibitem{std3gpp38901}
\emph{5G; Study on channel model for frequencies from 0.5 to 100 GHz}, 3GPP Std. TR 38.901, Rev. 16.1.0, 2020.

\bibitem{std3gpp38211}
\emph{5G; NR; Physical channels and modulation}, 3GPP Std. TS 38.211, Rev. 16.2.0, 2020.

\bibitem{std3gpp38212}
\emph{5G; NR; Multiplexing and channel coding}, 3GPP Std. TS 38.212, Rev. 16.2.0, 2020.

\bibitem{Dong2007}
X.~Dong, W.-S. Lu, and A.~C. Soong, ``{Linear Interpolation in Pilot Symbol Assisted Channel Estimation for OFDM},'' \emph{{IEEE} Trans. Wireless Commun.}, vol.~6, no.~5, pp. 1910--1920, 2007.

\end{thebibliography}

\end{document}